\documentclass[twocolumn,english,aps,superscriptaddress,prb,floatfix,longbibliography,nofootinbib]{revtex4-1}
\usepackage[latin9]{inputenc}
\usepackage{color}
\usepackage{babel}
\usepackage{amsmath}
\usepackage{amssymb}
\usepackage{mathtools}
\usepackage{graphicx}
\usepackage{wasysym}
\usepackage{esint}
\usepackage[unicode=true,pdfusetitle,
 bookmarks=true,bookmarksnumbered=false,bookmarksopen=false,
 breaklinks=true,pdfborder={0 0 0},pdfborderstyle={},backref=false,colorlinks=true]
 {hyperref}
\hypersetup{
 linkcolor=blue,citecolor=blue,urlcolor=blue}

\makeatletter

\graphicspath{{./}{./figs/}}

\providecommand{\tabularnewline}{\\}

\usepackage[dvipsnames]{xcolor}
\usepackage{babel}

\newcommand{\cpar} {\chi_{\parallel}}
\newcommand{\cperp} {\chi_{\perp}}
\newcommand{\tcwpar}{\Theta_{\parallel}^{{\rm CW}}}
\newcommand{\tcwperp}{\Theta_{\perp}^{{\rm CW}}}
\newcommand{\gpar} {g_{\parallel}}
\newcommand{\gperp} {g_{\perp}}

\newcommand{\lllangle}{\langle\!\langle\!\langle}
\newcommand{\rrrangle}{\rangle\!\rangle\!\rangle}

\newcommand{\nio}{Na$_2$IrO$_3$}

\newcommand{\aio}{$A_2$IrO$_3$}
\newcommand{\rucl}{$\alpha$-RuCl$_3$}

\newcommand{\Tc}{T_{\rm N}}
\newcommand{\Tu}{T_{\rm u}}
\newcommand{\Tl}{T_{\rm l}}

\newcommand{\Nr}{N_{\rm rl}}

\makeatother

\begin{document}
\title{Susceptibility anisotropy and its disorder evolution in models for Kitaev materials}

\author{Eric C. Andrade}
\affiliation{Instituto de F\'isica de S\~ao Carlos, Universidade de S\~ao Paulo,
C.P. 369, S\~ao Carlos, SP, 13560-970, Brazil}
\author{Lukas Janssen}
\affiliation{Institut f\"ur Theoretische Physik and W\"urzburg-Dresden Cluster
of Excellence ct.qmat, Technische Universit\"at Dresden, 01062 Dresden,
Germany}
\author{Matthias Vojta}
\affiliation{Institut f\"ur Theoretische Physik and W\"urzburg-Dresden Cluster
of Excellence ct.qmat, Technische Universit\"at Dresden, 01062 Dresden,
Germany}
\begin{abstract}
Mott insulators with strong spin-orbit coupling display a strongly anisotropic response to applied magnetic fields. This applies in particular to Kitaev materials, with \rucl\ and \nio\ representing two important examples. Both show a magnetically ordered zigzag state at low temperatures, and considerable effort has been devoted to properly modeling these systems in order to identify routes towards realizing a quantum spin liquid.
Here, we investigate the relevant Heisenberg-Kitaev-$\Gamma$ model primarily at elevated temperatures, focusing on the characteristic anisotropy between the in-plane and out-of-plane uniform susceptibility. For \rucl, we find that the experimentally observed anisotropy, including its temperature dependence, can be reproduced by combining a large off-diagonal $\Gamma_1$ coupling with a moderate $g$-factor anisotropy. Moreover, we study in detail the effect of magnetic dilution and provide predictions for the doping evolution of the temperature-dependent susceptibilities.
\end{abstract}
\date{\today}
\maketitle


\section{Introduction}

Spin-orbit coupling has moved center stage in the field of quantum magnetism primarily because it generates new states of matter, with spin ices and Kitaev spin liquids as prominent examples.\cite{bramwell01,kitaev06} Sizeable spin-orbit coupling also renders the magnetic-field response of magnets highly non-trivial, leading to strongly anisotropic magnetization processes, novel field-induced states that may feature complex spin textures, and associated quantum phase transitions.\cite{hkfield_review, majumder15, johnson15, chern17, rousochatzakis18, das19, chern20}

Following Kitaev's seminal paper \cite{kitaev06} on a solvable spin-liquid model on the honeycomb lattice and the subsequent proposal for realizing Kitaev interactions in particular exchange geometries,\cite{jackeli09,chaloupka10} a number of layered honeycomb magnets have been synthesized and investigated with aim to find spin-liquid phases. Among those materials, \aio\ ($A = \text{Na}, \text{Li}$) and \rucl\ have been studied in considerable detail:\cite{choi12,singh12,plumb14,sears15,baek17,wolter17,sears17,gass20,bachus20} All of them order antiferromagnetically at low temperature, but display a number of anomalies. Many of these anomalies have been attributed on phenomenological grounds to proximate or field-induced spin-liquid behavior, such as excitation continua in neutron scattering\cite{banerjee16,banerjee18} and the approximately half-quantized thermal Hall effect\cite{kasahara18b} observed in \rucl.
In the discussion of spin liquidity, randomness due to crystalline defects or substitutions is an additional relevant ingredient: The compound H$_3$LiIr$_2$O$_6$ has been suggested as a quantum spin-liquid candidate, but is likely heavily disordered.\cite{kitagawa18,knolle19a} For both \nio\ and \rucl, magnetic dilution has been proposed as a route to suppress bulk magnetic order, and $\alpha$-(Ru$_{1-x}$Ir$_x)$Cl$_3$ for $x\geq0.22$ has been argued to show spin-liquid-like signatures.\cite{kelley17,baek20}

Remarkably, there is still no  established consensus about the proper microscopic modeling of the various Kitaev materials, because (i) the parameter space spanned by all symmetry-allowed interactions, including second-neighbor and third-neighbor couplings, is huge and (ii) results from ab-initio calculations tend to depend sensitively on structural input and computational details. Moreover, partially conflicting results have been reported based on fits of experimental data.

In this paper, we focus on modeling a particularly intriguing and, at the same time, simple magnetic property of Kitaev materials, namely the susceptibility anisotropy. This anisotropy is huge in \rucl, with the response for in-plane fields being much larger than that for out-of-plane fields, and a consistent and detailed modeling has not been achieved to our knowledge.
We employ both high-temperature expansions and classical Monte-Carlo (MC) simulations to extract the magnetic susceptibilities at elevated temperatures, and we use MC to track these as function of temperature across the magnetic ordering transition. At low temperatures, our results are shown to agree with those from spin-wave calculations for the same models, provided that domain averaging is properly taken into account. We argue that most of the measured anisotropy data in \rucl\ are consistent with the presence of a sizeable symmetric off-diagonal $\Gamma_1$ interaction combined with a moderate $g$-factor anisotropy. By contrast, the $\Gamma_1$ interaction is small in \nio.
To describe diluted Kitaev materials, we perform MC simulations for the spin models with a fixed number of randomly distributed vacancies and extract the magnetic properties as function of dilution level. We show that the low-temperature susceptibilities are strongly enhanced, displaying Curie-like tails; in the presence of a large $\Gamma_1$ interaction this applies in particular to the in-plane susceptibility. We make concrete predictions for the doping evolution of the anisotropy at different temperatures.

The remainder of the paper is organized as follows:
Section~\ref{sec:experiments} provides a quick overview of published experimental results concerning magnetic anisotropies in \rucl\ and \nio\ and their diluted versions.
In Sec.~\ref{sec:model}, we introduce the relevant microscopic models and simulation techniques. Section~\ref{sec:clean} then shows numerical results for the clean, i.e., undiluted case. Section~\ref{sec:dil} is devoted to the magnetically diluted systems.
A concluding discussion closes the paper.
The effects of the so-called $\Gamma_1'$ coupling is discussed in an appendix.


\section{\label{sec:experiments} Review of experimental status}

We focus on the honeycomb Mott insulators \rucl\ and \nio, for which a number of experimental results are available.\cite{hkfield_review,takagi19} Both compounds display a low-temperature transition towards an antiferromagnetically ordered state of the zigzag type.
In both cases, diluted sister compounds have been possible to synthesize by substituting magnetic Ru$^{3+}$ and Ir$^{4+}$ by nonmagnetic Ir$^{3+}$ (Refs.~\onlinecite{kelley17, do18, do20}) and Ti$^{4+}$ (Ref.~\onlinecite{manni14}), respectively.


\subsection{\rucl}

In \rucl, the out-of-plane susceptibility $\cperp$ for fields along the direction perpendicular to the honeycomb plane is significantly smaller than the in-plane susceptibility $\cpar$.\cite{sears15, majumder15, kubota15, inplanefoot}
Furthermore, this anisotropy is strongly temperature dependent:\cite{kelley18a} The ratio between in-plane and out-of-plane susceptibility $\cpar/\cperp$ is around 1.5 at room temperature and increases up to a maximum of around 9--10 close to the N\'eel temperature $\Tc \simeq 7$\,K in the most recent samples. The anisotropy ratio again decreases in the ordered phase and levels off at around 5--8 at the lowest available temperatures.\cite{kelley18b}
Note that these values are sample dependent, likely as a consequence of stacking faults:\cite{cao16} Earlier samples with two transitions at 8\,K and 14\,K exhibit a significantly less pronounced anisotropy;\cite{majumder15, kubota15} for instance, in the ordered phase, the anisotropy ratio in these early samples may be up to 50\%\ smaller than in the more recent single-transition samples. Microscopically, this may arise either from $\Gamma_1$ (and possibly $\Gamma_1'$) interactions modified by stacking faults, or by anisotropic inter-layer interactions being different for different stackings.\cite{hk3d}

The anisotropic susceptibility leads to a corresponding anisotropy in the Curie-Weiss temperatures for the different field directions. They have been estimated as $\tcwpar \simeq +(\text{30--60})$\,K for in-plane fields and $\tcwperp \simeq -(\text{150--200})$\,K for out-of-plane fields, depending on the sample and the particular fitting range.\cite{sears15, majumder15, kelley18a}


\subsection{$\alpha$-(Ru$_{1-x}$Ir$_x)$Cl$_3$}

Upon substituting Ru$^{3+}$ with nonmagnetic Ir$^{3+}$, current samples with $0<x\lesssim0.1$ show again two transition temperatures, indicating the presence of stacking faults.\cite{kelley17} Increasing the doping concentration $x$ suppresses both N\'eel temperatures. The magnetic anisotropy shows a peculiar $x$ dependence: The in-plane susceptibility $\cpar$ increases with $x$ at low temperatures, but decreases with $x$ at high temperatures. The out-of-plane susceptibility $\cperp$, on the other hand, displays at high temperatures the same decreasing trend with $x$ as $\cpar$, but exhibits at low temperatures a non-monotonic behavior as function of $x$: \cite{do18} $\cperp$ increases steeply with $x$ for small doping, decreases in an intermediate regime up to $x=0.1$, and increases again for $x>0.1$. The strong increase of $\cperp$ for small doping leads to a reduced anisotropy at low temperatures.

Similarly, for small doping, the magnitude of the Curie-Weiss temperature for out-of-plane fields strongly decreases with increasing $x$, while the in-plane Curie-Weiss temperature only moderately decreases.\cite{do18}


\subsection{\nio}

\nio\ displays a weaker and qualitatively different susceptibility anisotropy as compared to \rucl. For all measured temperatures, the in-plane susceptibility $\cpar$ is significantly smaller than the out-of-plane susceptibility $\cperp$, with an anisotropy ratio $\cpar/\cperp$ of around $0.6$ at room temperature, which decreases to around $0.4$ at low temperatures, with a weak dip at the N\'eel temperature $\Tc \simeq 15$\,K.\cite{singh10,choi12,niosuscfoot}
The Curie-Weiss temperatures for the two different directions have been estimated as $\tcwpar \simeq -(\text{150--200})$\,K and $\tcwperp \simeq -(\text{30--50})$\,K.\cite{manni14b}


\subsection{Na$_2$(Ir$_{1-x}$Ti$_x$)O$_3$}

Upon substituting Ir$^{4+}$ with nonmagnetic Ti$^{4+}$, a hysteresis between field-cooled and zero-field-cooled susceptibilities is found at low temperatures, characteristic of spin-glass behavior.\cite{manni14} The freezing temperature decreases with increasing $x$; at the same time, the low-temperature susceptibility steeply increases. The high-temperature part can be fitted to Curie-Weiss behavior, with the magnitude of the Curie-Weiss temperature  decreasing substantially with increasing $x$. The anisotropy of the susceptibility has not been studied to our knowledge.


\section{\label{sec:model}Model and Monte-Carlo simulation}

\begin{table*}
\caption{\label{tab:parset}Minimal spin models for \rucl\ and \nio\ used in this paper and resulting N\'eel temperatures $\Tc$ and Curie-Weiss temperatures $\tcwpar$ and $\tcwperp$. The N\'eel temperatures have been estimated from the MC simulations by replacing $S^2 \to S(S+1)$ with $S=1/2$, see text. With this replacement, the Curie-Weiss temperatures obtained from fits of the high-temperature MC data agree with the results from the high-temperature expansion.
}
\begin{tabular*}{\linewidth}{@{\extracolsep\fill}ccccccccc}
\hline\hline
Model & Material & $J_{1}$ [meV] & $K_1$ [meV] & $\Gamma_1$ [meV] & $J_{3}$ [meV] & $\Tc$ [K] & $\tcwpar$ [K] & $\tcwperp$ [K] \tabularnewline
\hline
$1$ & \rucl\ & $-0.5$ & $-5.0$ & $2.5$ & $0.5$ & 9 & 22 & 0\tabularnewline
$2$ & \rucl\ & $-1.7$ & $-6.6$ & $6.6$ & $2.7$ & 34 & 30 & -28\tabularnewline
$3$ & \nio\ & $0$ & $-17$ & $0$ & $6.8$ & 48 & -10 & -10 \tabularnewline
\hline\hline
\end{tabular*}
\end{table*}

As a minimal model to understand the novel physics of the Kitaev materials,
we consider the extended Heisenberg-Kitaev-$\Gamma$ (HK$\Gamma$)
model \citep{chaloupka10,rau14}
\begin{align}
\mathcal{H} & =\sum_{\left\langle ij\right\rangle _{\gamma}}\left[J_{1}\vec{S}_{i}\cdot\vec{S}_{j}+K_1 S_{i}^{\gamma}S_{j}^{\gamma}+\Gamma_1 \left(S_{i}^{\alpha}S_{j}^{\beta}+S_{i}^{\beta}S_{j}^{\alpha}\right)\right]\nonumber \\
 &\quad  +J_{3}\sum_{\lllangle ij \rrrangle}\vec{S}_{i}\cdot\vec{S}_{j}-g_{\alpha}h_{\alpha}\sum_{i,\alpha}S_{i}^{\alpha}.
\label{eq:hkg}
\end{align}
Here, $J_{1,3}$ is the first-neighbor and third-neighbor Heisenberg coupling, and $K_1$ and $\Gamma_1$ are the first-neighbor Kitaev and off-diagonal symmetric couplings, respectively. $\left\langle ij\right\rangle _{\gamma}$ denotes the first-neighbor $\gamma$ bond, with $\gamma=x,y,z$, and $\lllangle ij\rrrangle$ denotes third-neighbors along opposite points of the same hexagon. $(\alpha,\beta,\gamma )= (y,z,x)$, $(z,x,y)$, and $(x,y,z)$ for the $x$, $y$, and $z$ bonds, respectively. The magnetic field is $\vec{h}:=\mu_\mathrm{B}\mu_{0}\vec{H}$, with $\mu_\mathrm{B}$ the Bohr magneton, and $g_{\alpha}$ are the diagonal components of the effective $g$ tensor. The symmetry-allowed nearest-neighbor couplings also include an additional $\Gamma_1'$ term which, however, is believed to be small. We defer its discussion to the appendix. Finally, we note that we neglect distortions which would spoil the $C_3^{\ast}$ symmetry of combined spin and lattice rotations in Eq.~\eqref{eq:hkg}. For \rucl\ such distortions are absent in the rhombohedral $R\bar{3}$ structure which is most likely realized below the structural transition at $100-150~$K.\cite{kubota15, glamazda17, reschke17}.

A large number of different parameter sets have been suggested to be relevant for the Kitaev materials \rucl\ and \nio, and we refer the reader to Ref.~\onlinecite{hk_field2} for a (partial) overview and discussion. In this paper, we shall employ three different minimal models displayed in Table~\ref{tab:parset}: Models 1 and 2 feature both large Kitaev and $\Gamma_1$ couplings and have been proposed to describe \rucl.\citep{rau14, hk_field2, rau14b, winter16, ran17, winter17b} By contrast, Model 3 has vanishing $\Gamma_1$ and has been suggested as a simple model for \nio.\citep{winter16, sizyuk14,sizyuk16,katukuri14}

We simulate Eq.~\eqref{eq:hkg} using classical MC simulations
on lattices of linear size $L$ and periodic boundary conditions.
Our spins are then replaced by classical vectors of fixed length $S$.
The honeycomb lattice is spanned by the primitive lattice vectors
$\vec{a}_{1\left(2\right)}=\left(3/2,\pm\sqrt{3}/2\right)$, with
each unit cell containing two sites, and thus $N=2L^{2}$, where $N$
is the total number of sites. Depletion is simulated by randomly removing
a fraction $x$ of spins, with $x$ varying between 5\% and 30\%,
with the total number of spins $N_{\mathrm{s}}=\left(1-x\right) N$.
We perform equilibrium MC simulations using single-site updates with
a combination of the heat-bath and microcanonical (or over-relaxation)
methods. For simulations at high temperatures, the simulations reach
equilibrium quickly and tens of thousands of MC steps per spin are
sufficient to evaluate the thermal averages. Disorder averages are
taken over $\Nr$ samples, with $\Nr\sim100$.


\section{Results for the clean limit}
\label{sec:clean}

\subsection{Ordering temperature}

The classical limit of the nearest-neighbor Heisenberg-Kitaev model for $\Gamma_1=J_{3}=0$
and $\vec{h}=0$ was previously investigated in Refs.~\onlinecite{price12,price13}.
The results in Refs.~\onlinecite{price12,price13} show two thermal transitions
upon cooling from the high-temperature paramagnetic phase to the ordered
zigzag phase. At $\Tu$, the system enters a critical phase with power-law
spin correlations, and at $\Tl<\Tu$ the zigzag state is reached.
This unusual behavior manifests itself, for instance, in a maximum
of the specific heat, or magnetic susceptibility, above both $\Tl$
and $\Tu$. There is a well-defined crossing point in $\xi(T)/L$ at $\Tl$ for different system sizes, where $\xi\left(T\right)$ is the zigzag correlation length, below which long-range order appears.
The existence of the critical intermediate phase has been related to the behavior of the two-dimensional six-state clock model,\cite{jose77} given that the ordered state of the Heisenberg-Kitaev model is sixfold degenerate.
In the presence of a weak interlayer exchange coupling, the critical phase disappears and $\Tl$ coincides with $\Tc$, the ordering temperature.\cite{hk_depl14}

\begin{figure}
\textbf{Model 1}
\vspace{2mm}

\begin{centering}
\includegraphics[width=0.5\columnwidth]{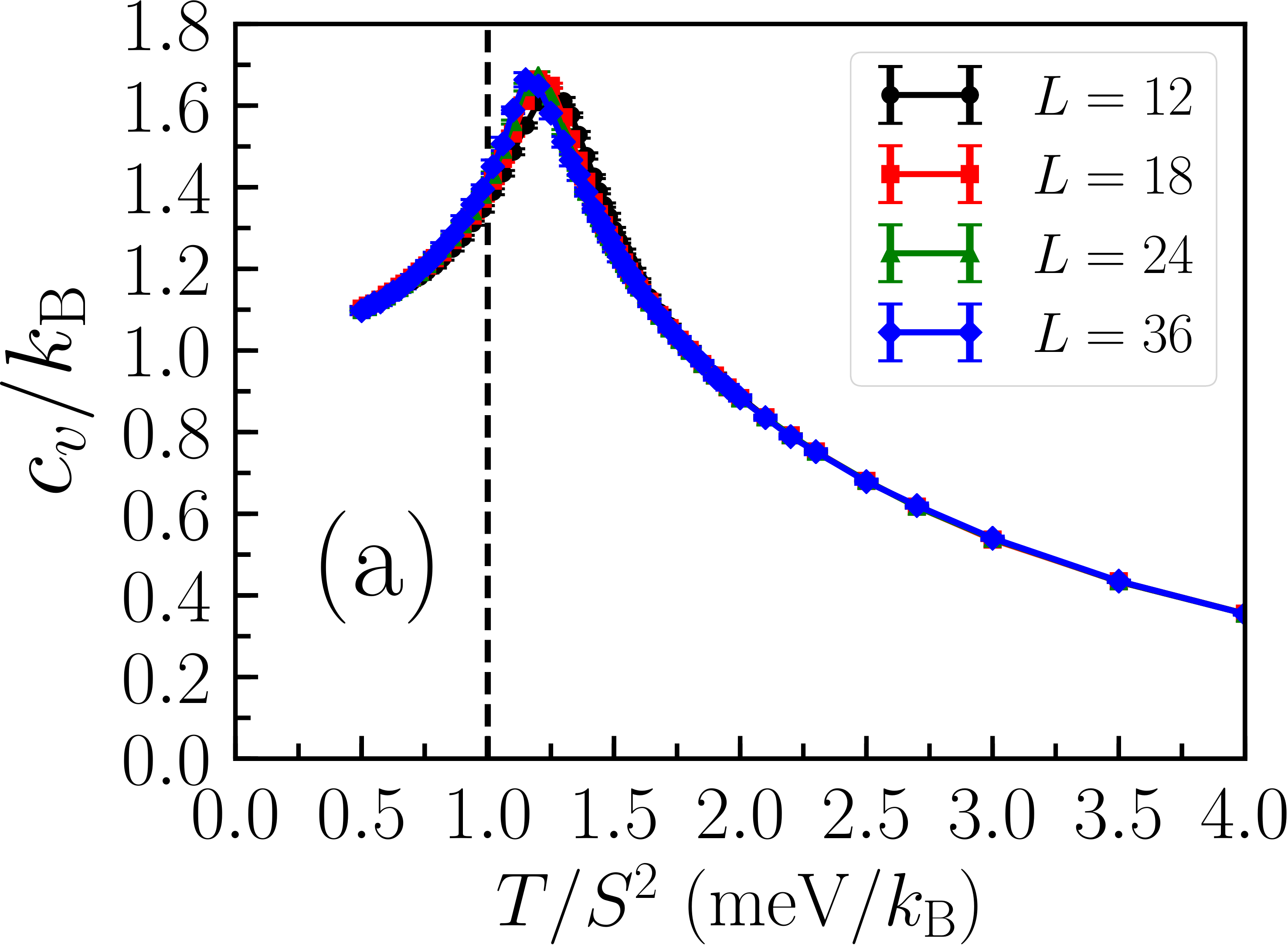}$\,$\includegraphics[width=0.5\columnwidth]{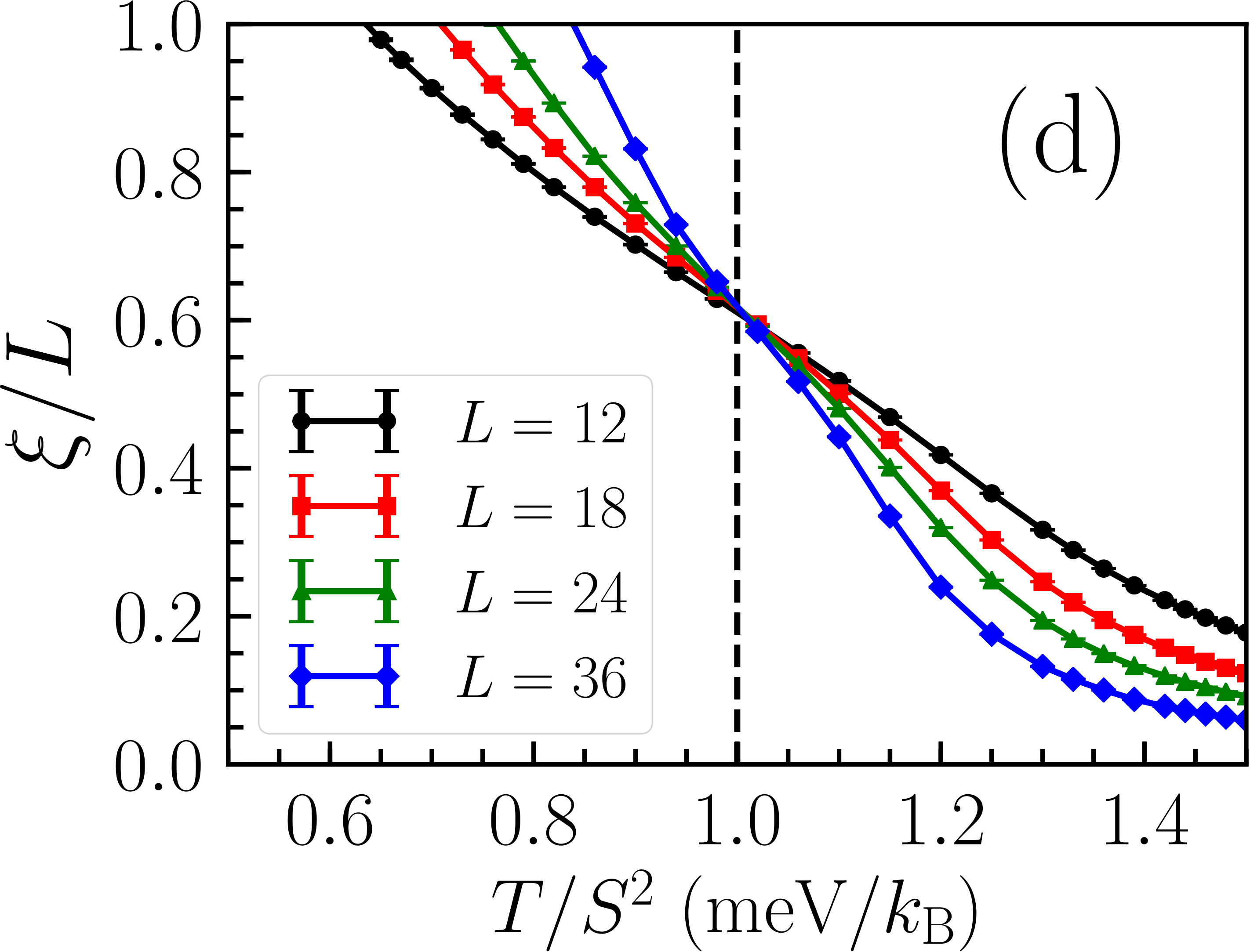}
\par\end{centering}

\textbf{Model 2}
\vspace{2mm}

\begin{centering}
\includegraphics[width=0.5\columnwidth]{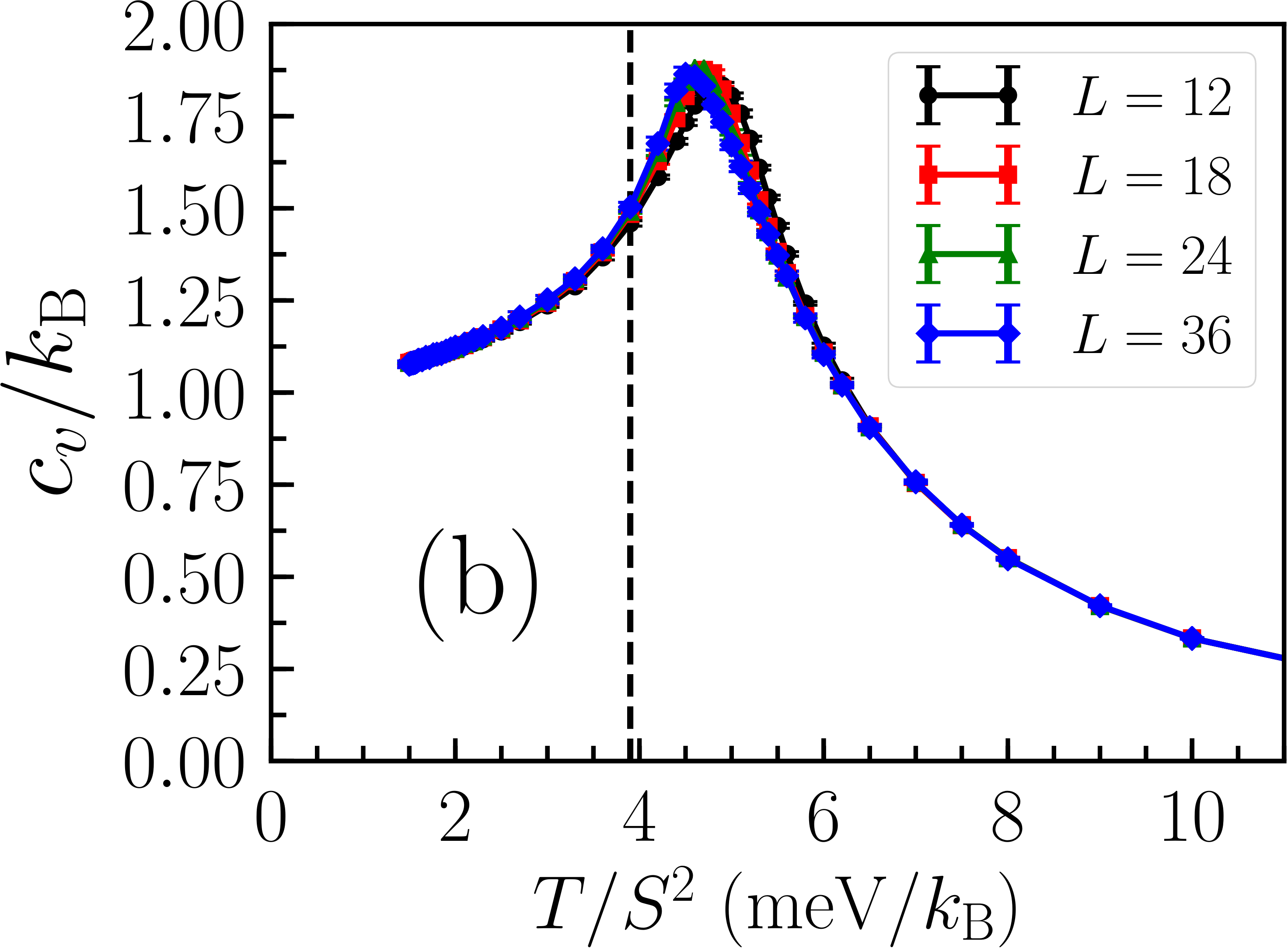}$\;$\includegraphics[width=0.5\columnwidth]{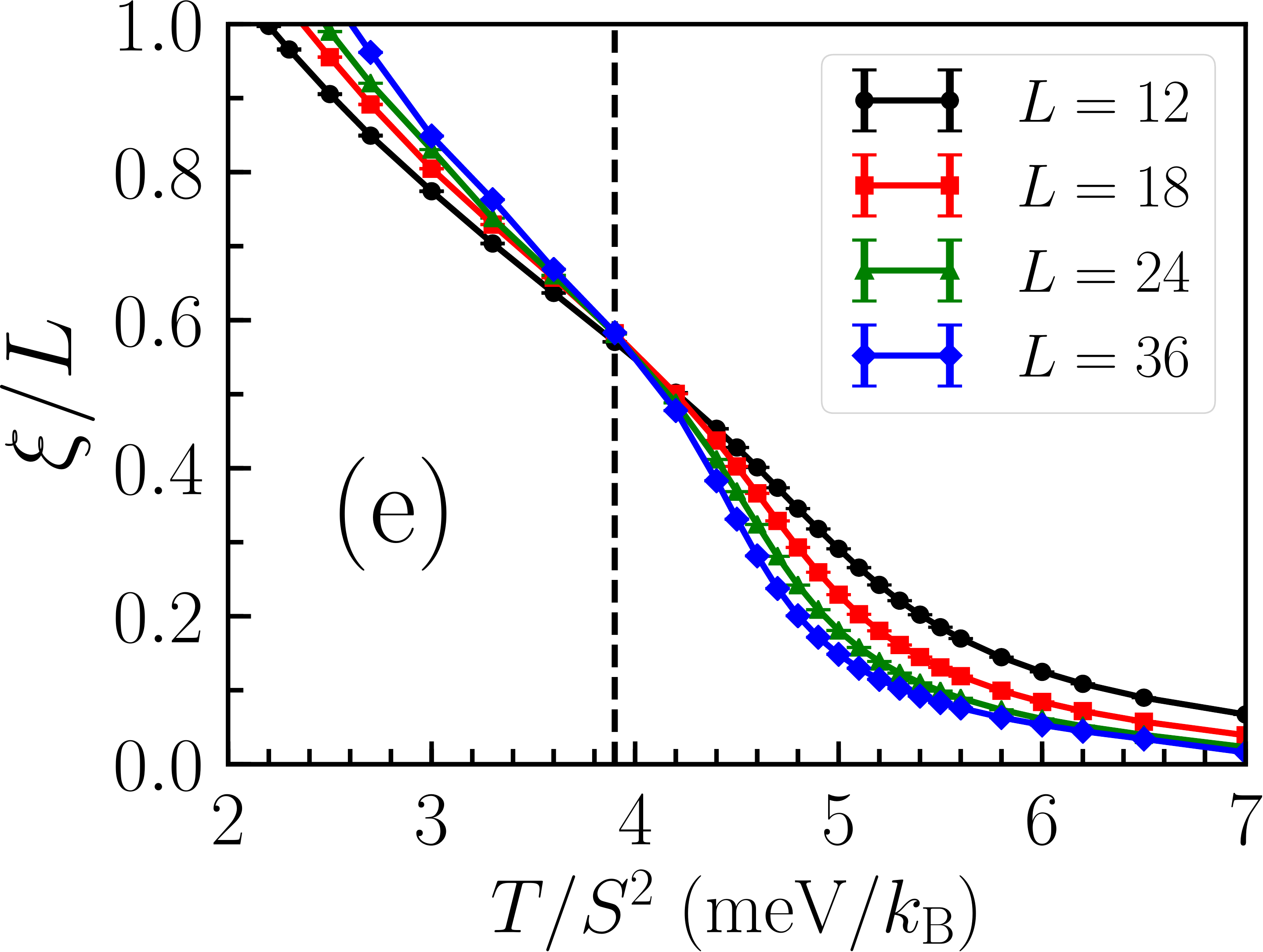}
\par\end{centering}

\textbf{Model 3}
\vspace{2mm}

\begin{centering}
\includegraphics[width=0.5\columnwidth]{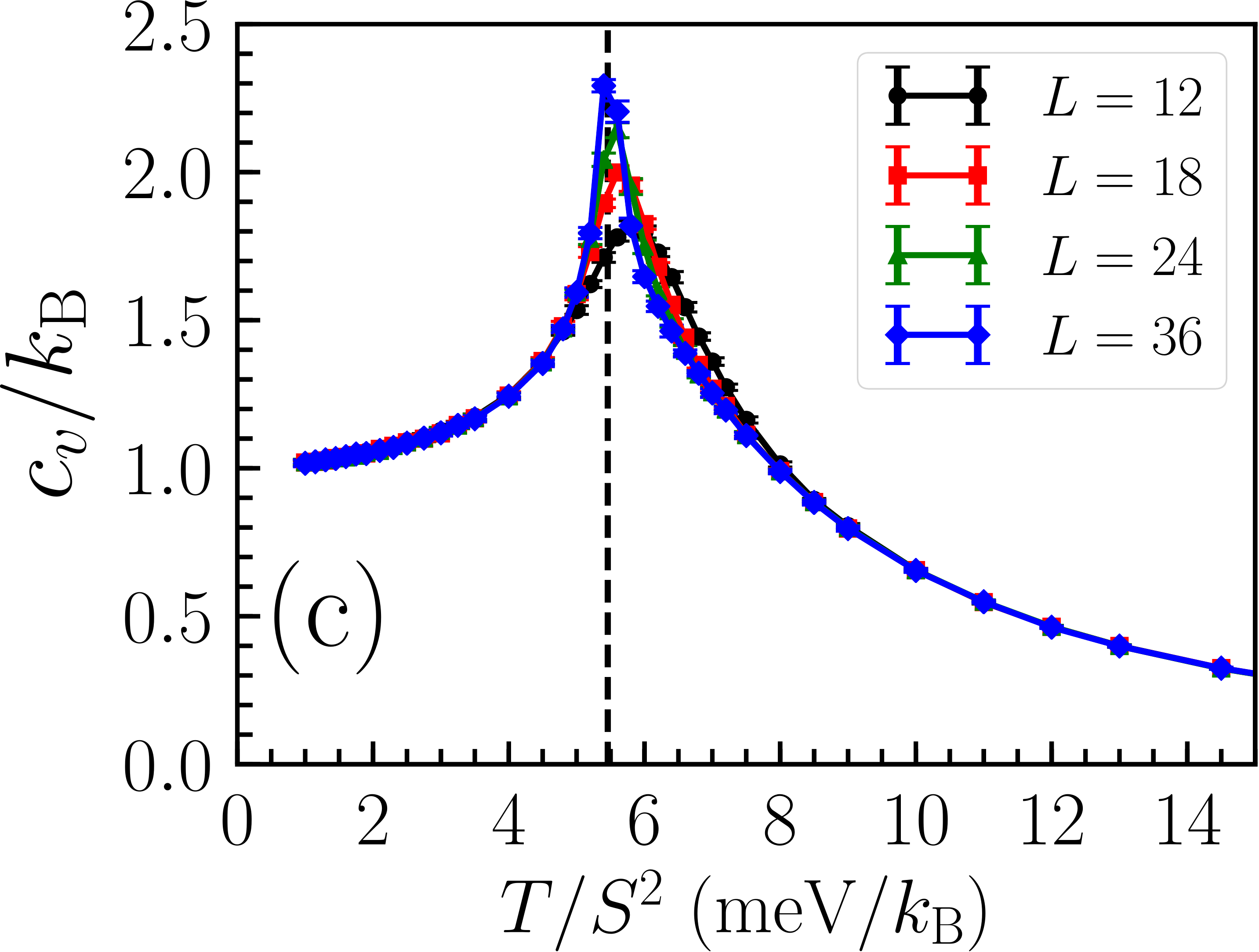}$\;$\includegraphics[width=0.5\columnwidth]{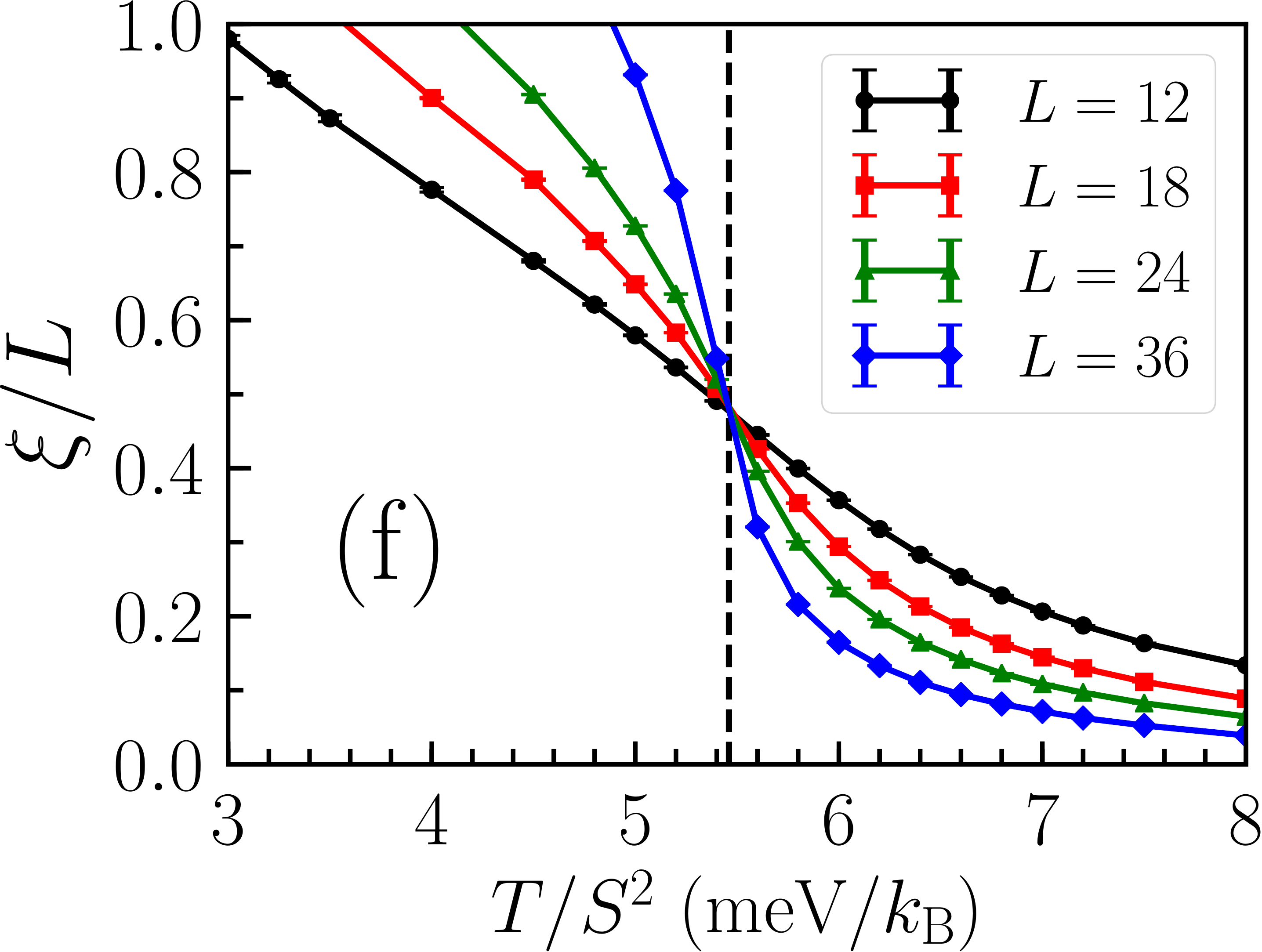}
\par\end{centering}
\centering{}\caption{\label{fig:clean}Specific heat $c_v$ (a--c) and zigzag correlation length $\xi$ (d--f) as function of temperature $T$ for Model 1 (a,d), Model 2 (b,e), and Model 3 (c,f) and different linear lattice sizes $L$.
The vertical dashed line marks the position of the ordering temperature $\Tc$.
}
\end{figure}

The $\Gamma_1$ term does not break the $C_{3}^{*}$ symmetry and hence preserves the degeneracy of the six zigzag domains. However, the accidental continuous degeneracy of the moment direction (at $T=0$) is lifted for $\Gamma_1>0$.\cite{chaloupka15,chaloupka16,sizyuk16,hk_field2} As shown in Figs.~\ref{fig:clean}(a) and (b), the specific heat for the HK$\Gamma$ model displays a maximum with weak $L$ dependence above the transition temperature. This is analogous to the behavior of the Heisenberg-Kitaev model,\cite{price12,price13} suggesting that a critical phase above the ordering temperature is still present. Since the existence of this critical phase is restricted to the purely two-dimensional case, we will not investigate it further and simply assume that $\Tc=\Tl$.\cite{hk_depl14}

For the $K_1$-$J_{3}$ model, Fig.~\ref{fig:clean}(c), the specific-heat peak approximately coincides with the crossing point in $\xi(T)/L$, suggesting that the critical phase is absent in this model. This may
be due to the fact that the mechanism for the stabilization of zigzag
order is different in this case.\cite{hk_field2} In particular, there is a U(1)
degeneracy of the classical ground states for each zigzag propagation
direction, in contrast to the situation in the generic HK$\Gamma$ model.
If ones assumes a second-order transition and attempts finite-size scaling of the data in Figs.~\ref{fig:clean}(c) and (f), using either Ising or 3-state Potts critical exponents, one does not get a good data collapse. This may either suggest a weak first-order transition, similar to, e.g., the situation in the six-state Potts model,\cite{iino19} or the presence of a critical phase in a much reduced interval. Deciding between these scenarios is beyond the scope of this work.

To extract the ordering temperature $\Tc$, we adopt a pragmatic approach, which works well for moderate system sizes, and look for crossing points in the curve $\xi(T)/L$ for different system sizes, where $\xi\left(T\right)$ is the zigzag correlation length. The results are displayed in Figs.~\ref{fig:clean}(d), (e), and (f), from which we extract $k_\mathrm{B}\Tc/S^{2}=1.00(5)$ meV, $k_\mathrm{B}\Tc/S^{2}=3.9\left(1\right)$ meV, and $k_\mathrm{B}\Tc/S^{2}=5.46\left(6\right)$ meV, for the Models 1, 2, and 3, respectively.

We may contrast these ordering temperatures with the experimentally
measured N\'eel temperatures.
As we demonstrate explicitly when comparing our MC data with the results from the high-temperature expansion below, at high temperatures, quantum fluctuations amount to an effective energy rescaling, which can be incorporated by replacing $S^2 \to S(S+1)$.
This way, a simple estimate of the N\'eel temperatures for $S=1/2$ can then be obtained from the classical MC simulations, see Table~\ref{tab:parset}.
In the case of Model 1, which is assumed to be relevant to \rucl,\citep{winter17b,hk_field2} we find $\Tc\sim9$\,K for $S=1/2$, which is in the range of the experimentally reported values.\cite{sears15,majumder15,kubota15,banerjee16,banerjee18,baek17,wolter17,sears17,kelley18a}
A similar exercise for Models 2 and 3 produces $\Tc \sim 34$\,K and $\Tc \sim 48$\,K, respectively, which are both unrealistically high.\cite{singh10,singh12,choi12}

\subsection{Susceptibilties and Curie-Weiss temperatures}

\begin{figure}

\textbf{Model 1}
\vspace{2mm}

\begin{centering}
\includegraphics[width=0.5\columnwidth]{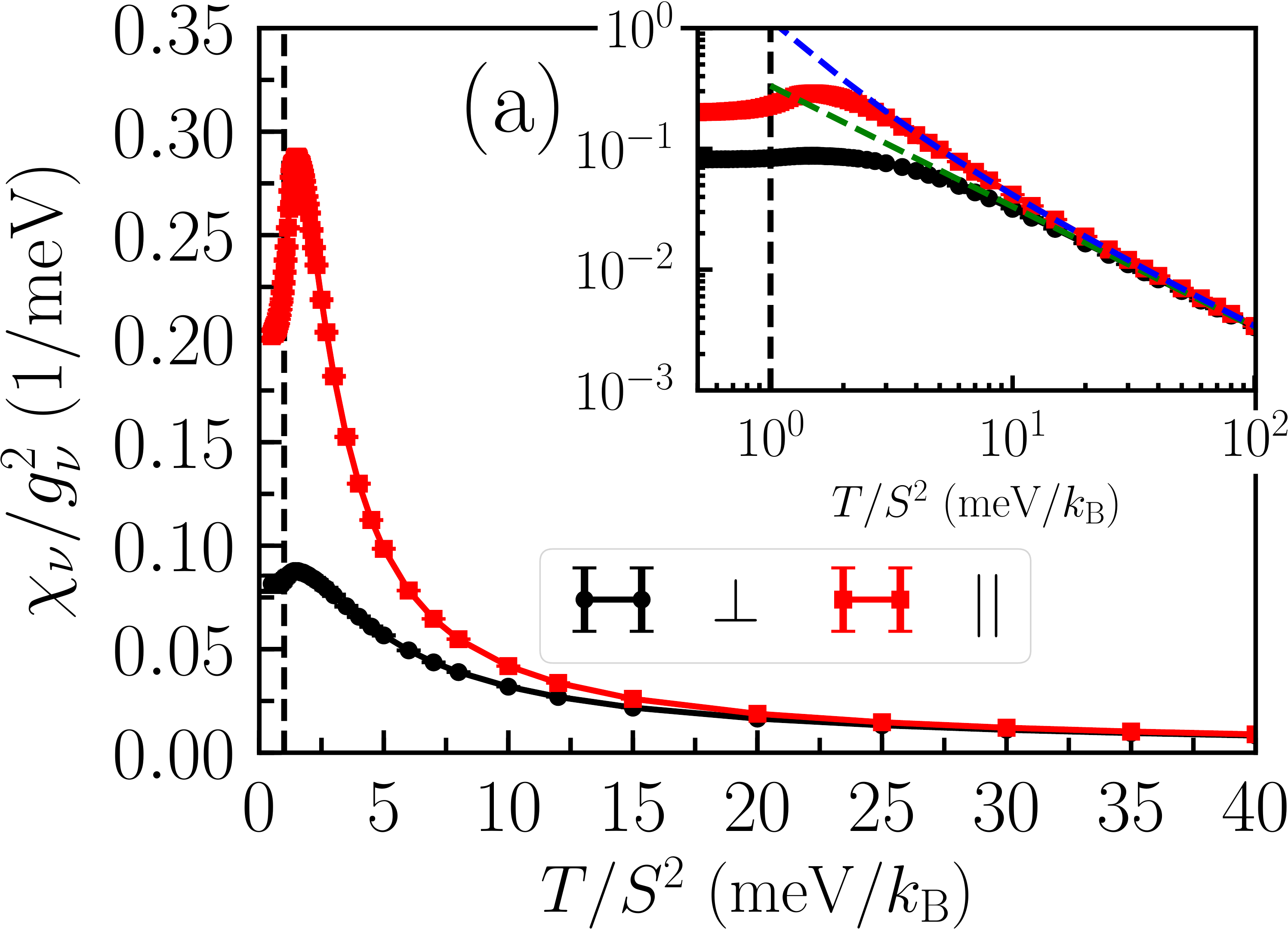}$\;$\includegraphics[width=0.47\columnwidth]{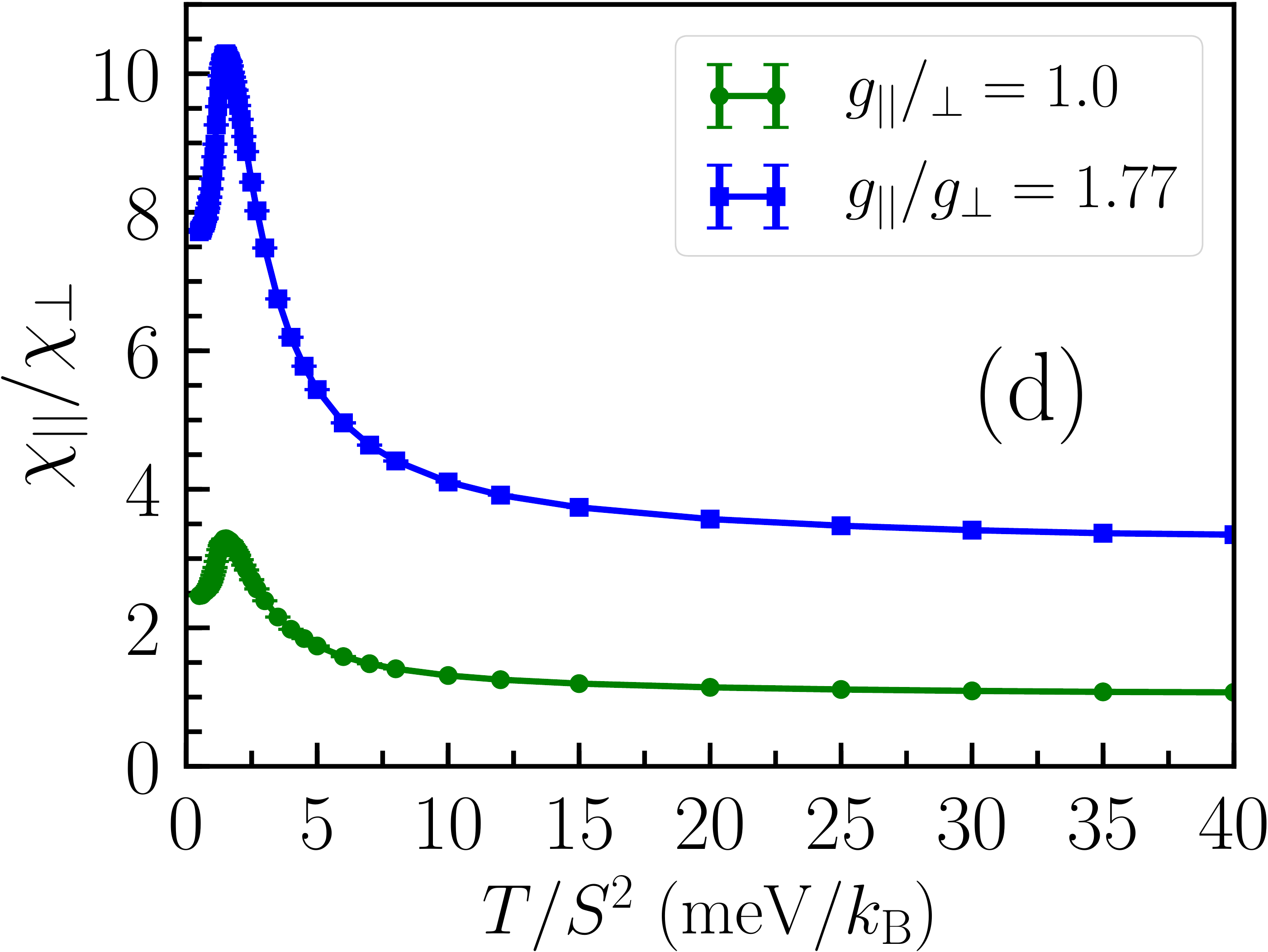}
\par\end{centering}

\textbf{Model 2}
\vspace{2mm}

\begin{centering}
\includegraphics[width=0.5\columnwidth]{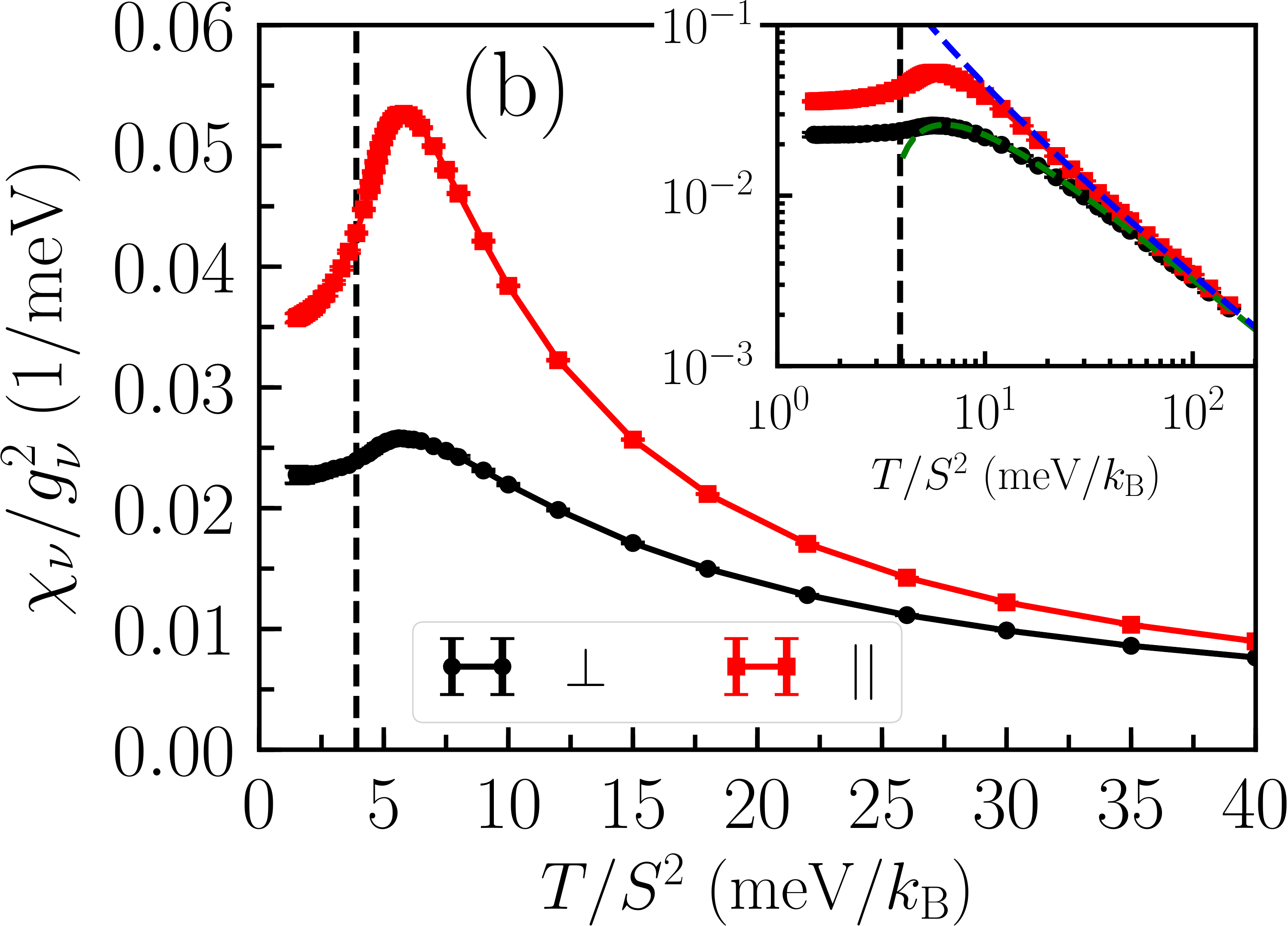}$\;$\includegraphics[width=0.47\columnwidth]{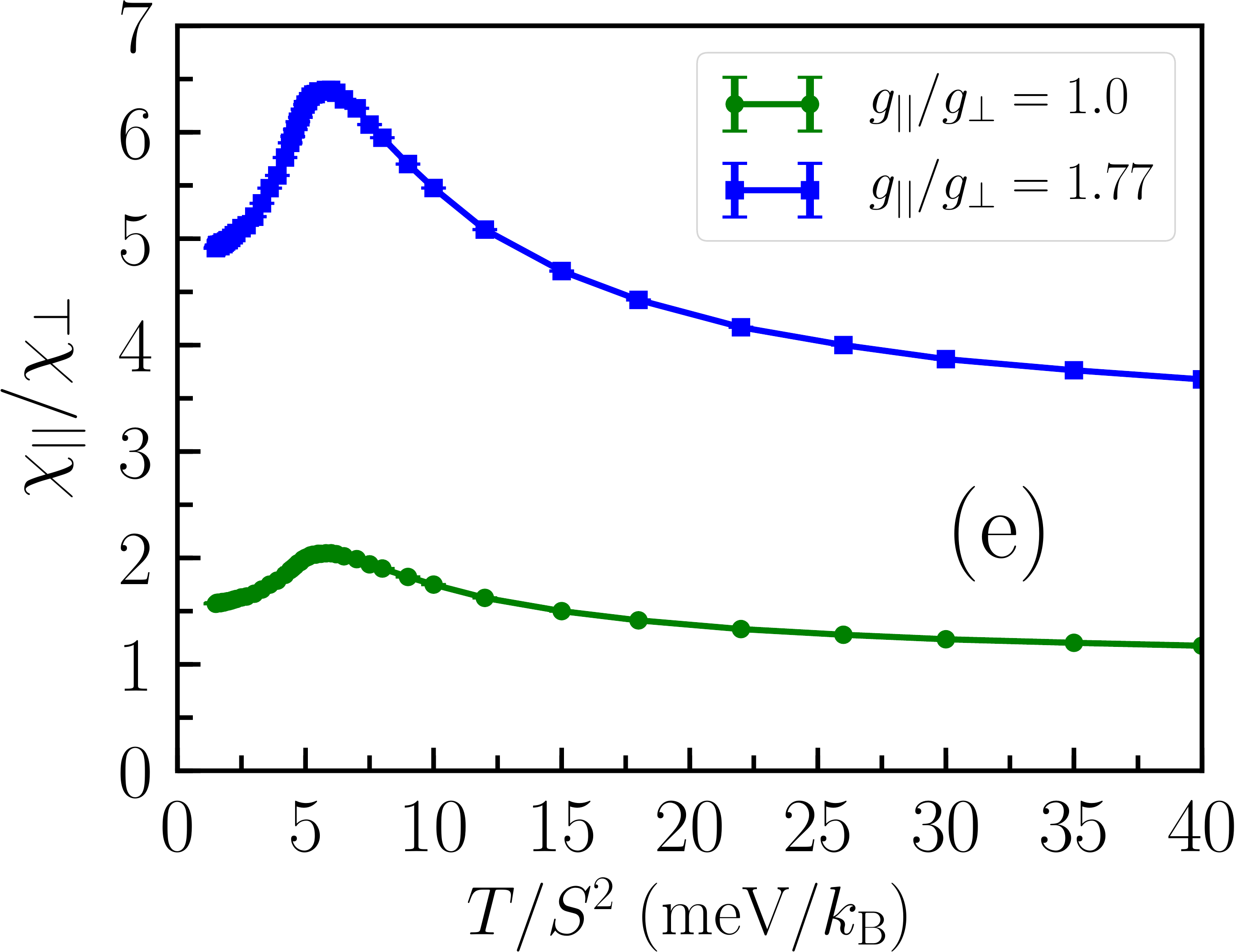}
\par\end{centering}

\textbf{Model 3}
\vspace{2mm}

\begin{centering}
\includegraphics[width=0.5\columnwidth]{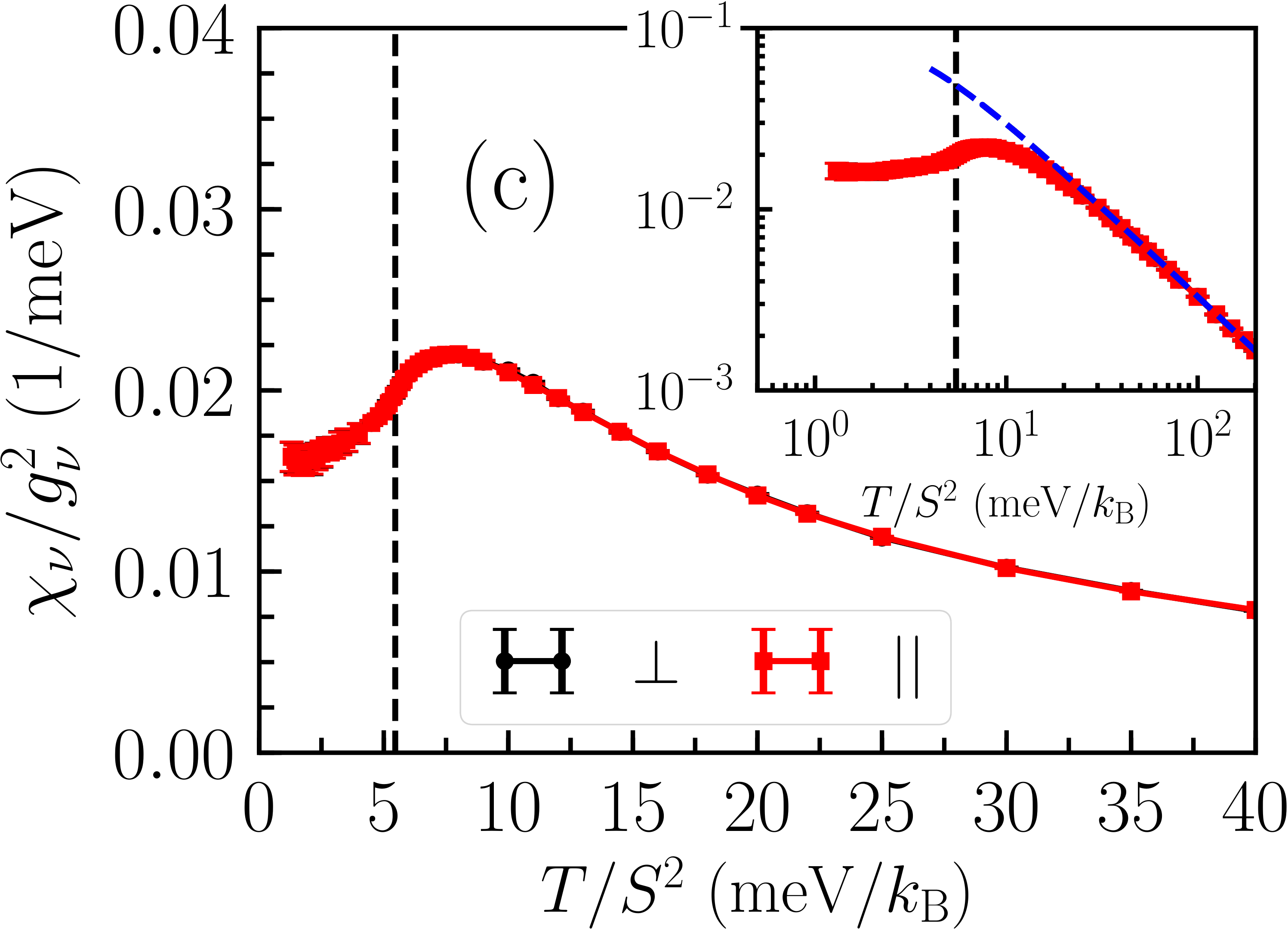}\includegraphics[width=0.485\columnwidth]{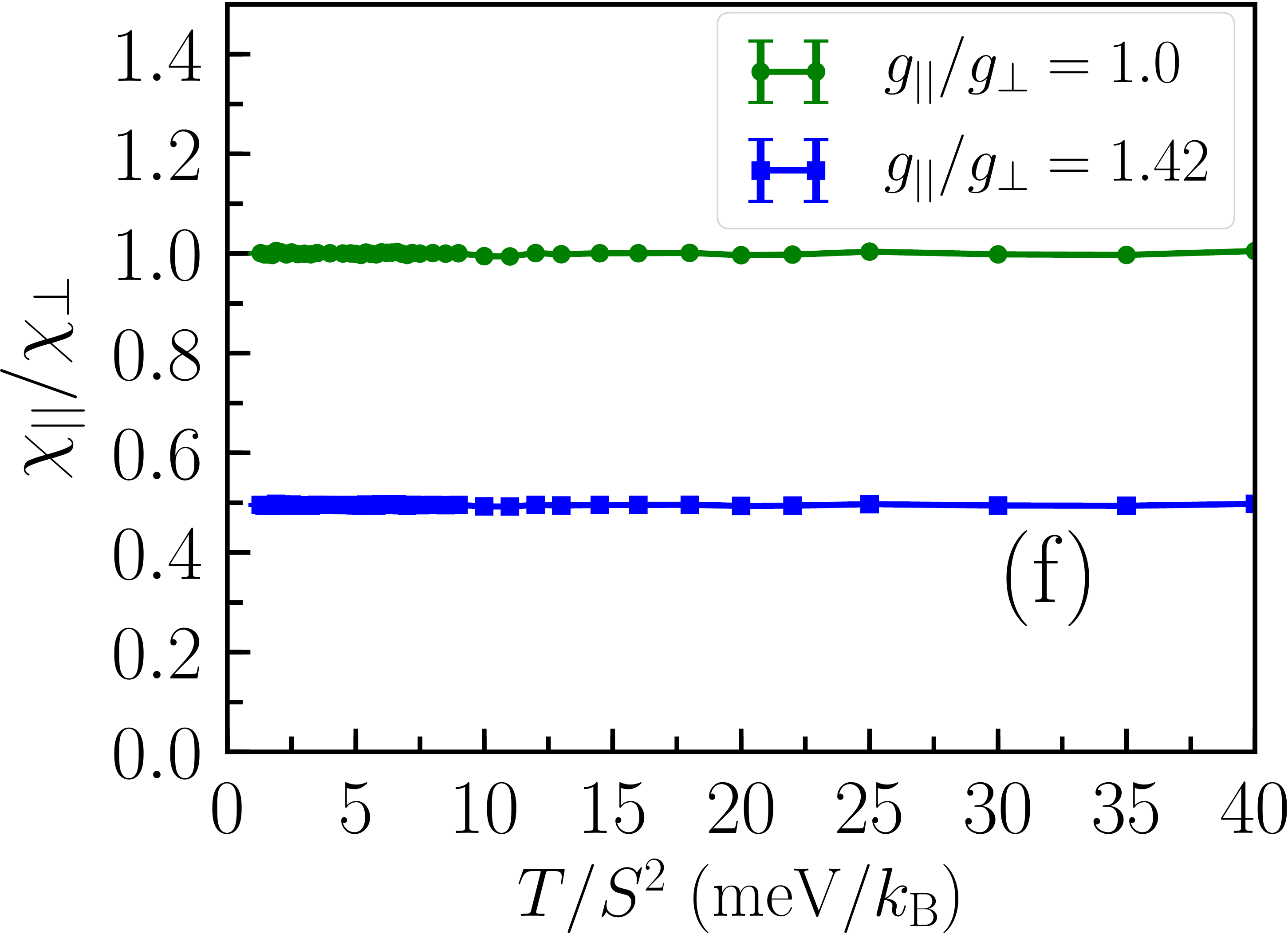}
\par\end{centering}

\caption{\label{fig:chi_clean} Out-of-plane susceptibility $\cperp$ (black) and in-plane susceptibilities $\cpar$ (red) for Model 1 (a), Model 2 (b), and Model 3 (c) as function of temperature $T$.
Susceptibility anisotropy ratio $\cpar/\cperp$ in units of $(\gpar/\gperp)^2$ for Model 1 (d), Model 2 (e), and Model 3 (f) as function of temperature $T$.
(a,b,c) MC results on an $L=24$ lattice in units of $\gpar^2$ and $\gperp^2$, displaying the $g$-factor independent intrinsic anisotropy arising from the $\Gamma_1$ interaction in (a) and (b). Inset: Comparison between MC results (full lines) and high-temperature series expansion (dashed lines) in a log-log scale, using $c = S^2$. The vertical dashed line marks the position of the ordering temperature $\Tc$.
(d,e,f) Susceptibility anisotropy ratio $\cpar/\cperp$ for two values of the ratio $(\gpar/\gperp)^2$ from the data in (a,b,c). We considered $\gpar=2.3$ and $\gperp=1.3$ (d,e) and $\gpar=1.9$ and $\gperp=2.7$ (f).
}
\end{figure}

From the MC data, we calculate the uniform magnetic susceptibilities as
\begin{eqnarray}
\chi_{\alpha\beta}\left(T\right)&=&\frac{\partial M_{\alpha}}{\partial h_{\beta}}=g_{\alpha}g_{\beta}\frac{N_\mathrm{s}}{T}\left\langle \sigma_{\alpha}\sigma_{\beta}\right\rangle,\label{eq:chi_uniform}
\end{eqnarray}
where $\left\langle \cdots\right\rangle $ denotes MC average, $M_{\alpha}=g_{\alpha}\left\langle \sigma_{\alpha}\right\rangle $
is the uniform magnetization, and $\sigma_{\alpha}=N_\mathrm{s}^{-1}\sum_{i}S_{i}^{\alpha}$,
with $\alpha,\beta \in \{x,y,z\}$. Because the ordered state carries no uniform
moment, we do not subtract the term $g_{\alpha}g_{\beta}\left\langle \sigma_{\alpha}\right\rangle \left\langle \sigma_{\beta}\right\rangle $,
which we expect to go to zero in the thermodynamic limit. In the linear-response limit of Eq.~\eqref{eq:chi_uniform}, the $g$ factors do not enter the MC simulations directly, appearing only as prefactors.

In \rucl, for example, the cubic axes point along nearest-neighbor
Ru-Cl bonds.\cite{hkfield_review} For convenience, we take the crystallographic $a$ and
$b$ axes along the directions $\left[11\bar{2}\right]$ and
$\left[\bar{1}10\right]$, respectively, defining the honeycomb
plane. The direction perpendicular to this plane is labeled as the
$\left[111\right]$ direction. These three directions form a
basis in which the susceptibility tensor is diagonal. Projecting
Eq.~\eqref{eq:chi_uniform} onto an in-plane direction then defines
the in-plane susceptibility
\begin{equation}
\cpar=\chi_{zz}-\chi_{xz}.\label{eq:cpar}
\end{equation}
Here, we have assumed that the three bond directions are equivalent, thus rendering only two independent components of the tensor susceptibility in Eq.~\eqref{eq:chi_uniform}: $\chi_{zz}$
and $\chi_{xz}$. Analogously, projecting the susceptibility tensor onto the $\left[111\right]$ direction perpendicular to the honeycomb layer gives the out-of-plane susceptibility
\begin{equation}
\cperp=\chi_{zz}+2\chi_{xz}.\label{eq:cperp}
\end{equation}

For an isotropic $g$ tensor, we obtain $\chi_{xz}\neq0$ only if $\Gamma_1\neq0$, and thus the anisotropy in the susceptibility can be traced back to the off-diagonal coupling $\Gamma_1$. In Ref.~\onlinecite{hk_field2} it was pointed out that the $\Gamma_1$ term alone, without any $g$-factor anisotropy, can lead to a ratio $\cpar/\cperp\sim3-4$ at low temperatures, i.e., deep in the ordered zigzag phase.
Note that these values correspond to the susceptibilities of the respective thermodynamically stable finite-field single-domain state. In \rucl\ such field-driven domain selection only happens for fields above 2\,T;\cite{banerjee18} at smaller fields, typically all zigzag domains contribute to the susceptibility, with relative weights that depend on the sample.\cite{kelley18b}

\subsubsection{High-temperature expansion}

To get started, it is useful to consider a high-temperature series expansion of the model in Eq.~\eqref{eq:hkg}, which we link both to experiments and to the MC simulations. The leading correction to the Curie susceptibility is given by\cite{rau14,kelley18a,singh17}
\begin{align}
\cpar & = \gpar^2 \frac{c}{3T}\left( 1+ \frac{\tcwpar}{T} \right) +\mathcal{O}(T^{-3}),\label{eq:cpar_highT}\\
\cperp & = \gperp^2 \frac{c}{3T}\left( 1 + \frac{\tcwperp}{T} \right) +\mathcal{O}(T^{-3}),\label{eq:cperp_highT}
\end{align}
with the constant
\begin{align}
c =
\begin{cases}
S(S+1) & \text{for quantum spins},\\
S^2 & \text{for classical spins},
\end{cases}
\end{align}
and where $\gpar$ ($\gperp$) is the $g$ factor along an in-plane (out-of-plane) direction. The asymptotic Curie-Weiss temperatures, defined as $\chi^{-1}_{\parallel,\perp}\left(T\right)\sim T - \Theta_{\parallel,\perp}^{\rm{CW}} + \mathcal{O}\left(T^{-1}\right)$, are given by
\begin{align}
\tcwpar & = -\frac{c}{3}\left[3\left(J_{1}+J_{3}\right)+K_1-\Gamma_1\right],\label{eq:cw_par}\\
\tcwperp & = -\frac{c}{3}\left[3\left(J_{1}+J_{3}\right)+K_1+2\Gamma_1\right].\label{eq:cw_perp}
\end{align}
These expressions explicitly show that an intrinsic anisotropy in the model can be traced back to the $\Gamma_1$ term.

\subsubsection{MC results}

In Figs.~\ref{fig:chi_clean}(a,b,c), we plot the out-of-plane and in-plane susceptibilities measured in the MC simulations together with the high-temperature results of Eqs.~\eqref{eq:cpar_highT} and \eqref{eq:cperp_highT}, using $c = S^2$. For Model 3, we obtain $\cpar=\cperp$, Fig.~\ref{fig:chi_clean}(c), since there is no intrinsic anisotropy for $\Gamma_1=0$. As in the specific-heat curves in Fig. \ref{fig:clean}, the maximum in the susceptibility curves (more visible in $\cpar$) occurs above $\Tc$. At very high $T$ we recover Curie's law, whereas the low-temperature susceptibilities become essentially constant as expected for an antiferromagnetic state. We find finite-size effects for this susceptibility curves to be small, and the $L=24$ result is representative of the
thermodynamic behavior.

Reassuringly, the susceptibility anisotropy found at the lowest temperatures in the MC simulation, Fig.~\ref{fig:chi_clean}(a), essentially agrees with that found for the same model from $T=0$ spin-wave calculations.\cite{hk_field2} For this comparison it is important to notice that the MC simulation, performed in the linear-response limit, averages over all zigzag domains, whereas the spin-wave calculation is performed for a particular domain. The results of both agree once domain averaging in the spin-wave calculation is properly performed.

\subsubsection{Comparison to experiments on \rucl}

For Models 1 and 2 in Table \ref{tab:parset}, which are thought to be relevant for \rucl, we find that $\cpar>\cperp$ for all temperatures we consider, with the maximum value of $(\cpar/\gpar^2)/(\cperp/\gperp^2)$ close to $3.3$ (Model 1) or $2.0$ (Model 2) in the paramagnetic regime just above $\Tc$, see Figs.~\ref{fig:chi_clean}(d,e). This is smaller than the values for $\cpar/\cperp$ observed in \rucl\ experiments, indicating  that either (i) the present models have to be significantly adapted\cite{kelley18a} or (ii) the observed magnetic anisotropy has further extrinsic contributions arising from a $g$-factor anisotropy. Since in particular Model 1 (or slight modifications thereof\cite{hk3d}) is found to describe well many properties of \rucl\ at low temperatures,\cite{winter17,hkfield_review} we consider it likely that anisotropies in the $g$ factor are present due to a trigonal crystal field and should be taken into account.
Further effects arising from a $\Gamma_1^{\prime}$ interaction are discussed in the appendix.

Previous work\cite{chaloupka16,yadav16,winter18} suggested that $\gperp \simeq 1.3$ and $\gpar \simeq 2.3$ at low $T$. In Figs.~\ref{fig:chi_clean}(d,e) we explore the effects of including these anisotropic $g$ factors, assuming that their temperature dependence is small. For Model 1 (Model 2), this gives $\cpar/\cperp \approx 3.3$ $(3.7)$ at $T \simeq 40 c\,\text{meV}/k_\mathrm{B} \simeq 350$\,K for $S=1/2$. Close to the transition to the zigzag phase, we obtain $\cpar/\cperp \approx 10$ $(6.4)$, whereas at low temperature $\cpar/\cperp \approx 7.5$ $(5)$. Overall, Model 1 provides a satisfactory description of the experimental trend, even if the anisotropy at room temperature is too large.
Part of this disagreement may be due to the fact that the single-transition samples---incidentally, those samples which agree quite well with Model 1 at low and intermediate temperatures---have a first-order structural transition between 100\,K and 150\,K, which has a sizeable effect on the out-of-plane susceptibility.\cite{do18, kelley18a} It is thus possible that the high-temperature phase requires a different modeling, especially of $\cperp$.
In the appendix, we demonstrate that inclusion of a small positive $\Gamma^{\prime}_1$ may further improve the overall agreement with the experimental data.

We note that a previous attempt\cite{kelley18a} of fitting \rucl\ model parameters from high-temperature susceptibilities using a nearly isotropic $g$ factor \cite{agrestini17} yielded large values for the $\Gamma_1$ coupling of around $30$\,meV.
Since the $g$ factor enters quadratically in the susceptibility, see Eq.~\eqref{eq:chi_uniform}, the susceptibility anisotropy is very sensitive to a $g$-factor anisotropy.
As shown above, with moderate $g$ anisotropy, the experimental data can be described with moderately large $\Gamma_1$.

For $T\apprge4\Tc$, the asymptotic high-temperature expansion describes well the MC data, see insets of Figs.~\ref{fig:chi_clean}(a,b,c). The resulting anisotropic Curie-Weiss temperatures are $\tcwpar=2.5c\,\text{meV}/{k_\mathrm{B}}$ and $\tcwperp=0$ (Model 1), $\tcwpar=3.4c\,\text{meV}/{k_\mathrm{B}}$ and $\tcwperp=-3.2c\,\text{meV}/{k_\mathrm{B}}$ (Model 2). While the signs of the Curie-Weiss temperatures (mostly) agree with the experimental results,\cite{sears15, majumder15, kelley18a} the absolute values for $S=1/2$ as displayed in Table~\ref{tab:parset} are too small, in particular in the case of $\tcwperp$. As with the high-temperature susceptibility anisotropy, this discrepancy may be tied to the fact that $\cperp$ is particularly sensitive to the first-order structural transition.

We can contrast these result with an effective Curie-Weiss temperature\cite{singh17} obtained by performing a linear fit of the inverse susceptibility in a finite temperature interval, as done in experimental analyses. We choose the temperature interval $20c\,\text{meV}/k_\mathrm{B} \apprle T\apprle40c\,\text{meV}/k_\mathrm{B}$, which corresponds to $175\,\mathrm{K} \apprle T\apprle 350\,\mathrm{K}$ for $S=1/2$, similar to the one employed in Ref.~\onlinecite{kelley18a}.
This way, we obtain $\tcwpar=2.1(1)c$\,meV and $\tcwperp=-0.3(1)c$\,meV for Model 1, whereas for Model 2, we get $\tcwpar=2.0(1) c$ meV and $\tcwperp=-4.5(1)c$\,meV.
Thus, the magnitude of the effective $\tcwpar$ ($\tcwperp$) decreases (increases) with decreasing temperature.
Note that the temperature dependence of the effective Curie-Weiss temperature is enhanced in Model 2, which can be traced back to the larger value of $\Tc$ in this case.

\subsubsection{Comparison to experiments on \nio}

Without any $g$-factor anisotropy, Model 3, which was proposed as a minimal model for \nio, does not show any anisotropy in the susceptibility data, Fig.~\ref{fig:chi_clean}(c). However, previous susceptibility measurements\cite{singh10} pointed to a significantly anisotropic $g$ factor, suggesting $\gperp \simeq 2.7$ and $\gpar \simeq 1.9$. Using these estimates, the resulting susceptibilities in Model 3 are shown in Fig.~\ref{fig:chi_clean}(f).
The corresponding ratio $\cpar/\cperp \simeq 0.5$ roughly agrees with the experiments in \nio,\cite{singh10} supporting the view that the $\Gamma_1$ term is likely small in this compound.


\section{\label{sec:dil}Effects of dilution}

We now move to the evolution of the magnetic anisotropy with magnetic dilution by randomly removing a fraction $x$ of spins. Because the HK$\Gamma$ model is frustrated, we expect the combination of randomness and frustration to generically lead to a spin-glass ground state.\cite{villain79,hk_depl14,manni14, pyro-rf,aharony78,imry-ma}
In strictly two spatial dimensions, the freezing temperature is zero.\citep{maiorano18} Studying the thermodynamic glass transition thus requires a detailed description of the interlayer coupling,\citep{hk3d} and we thus leave this for future work. Here, our focus will be exclusively on the disorder evolution of the uniform magnetic susceptibility and its anisotropy.

\subsection{Heisenberg limit and Curie tail}

\begin{figure}
\begin{centering}

\textbf{Heisenberg model}
\vspace{2mm}

\includegraphics[width=1.0\columnwidth]{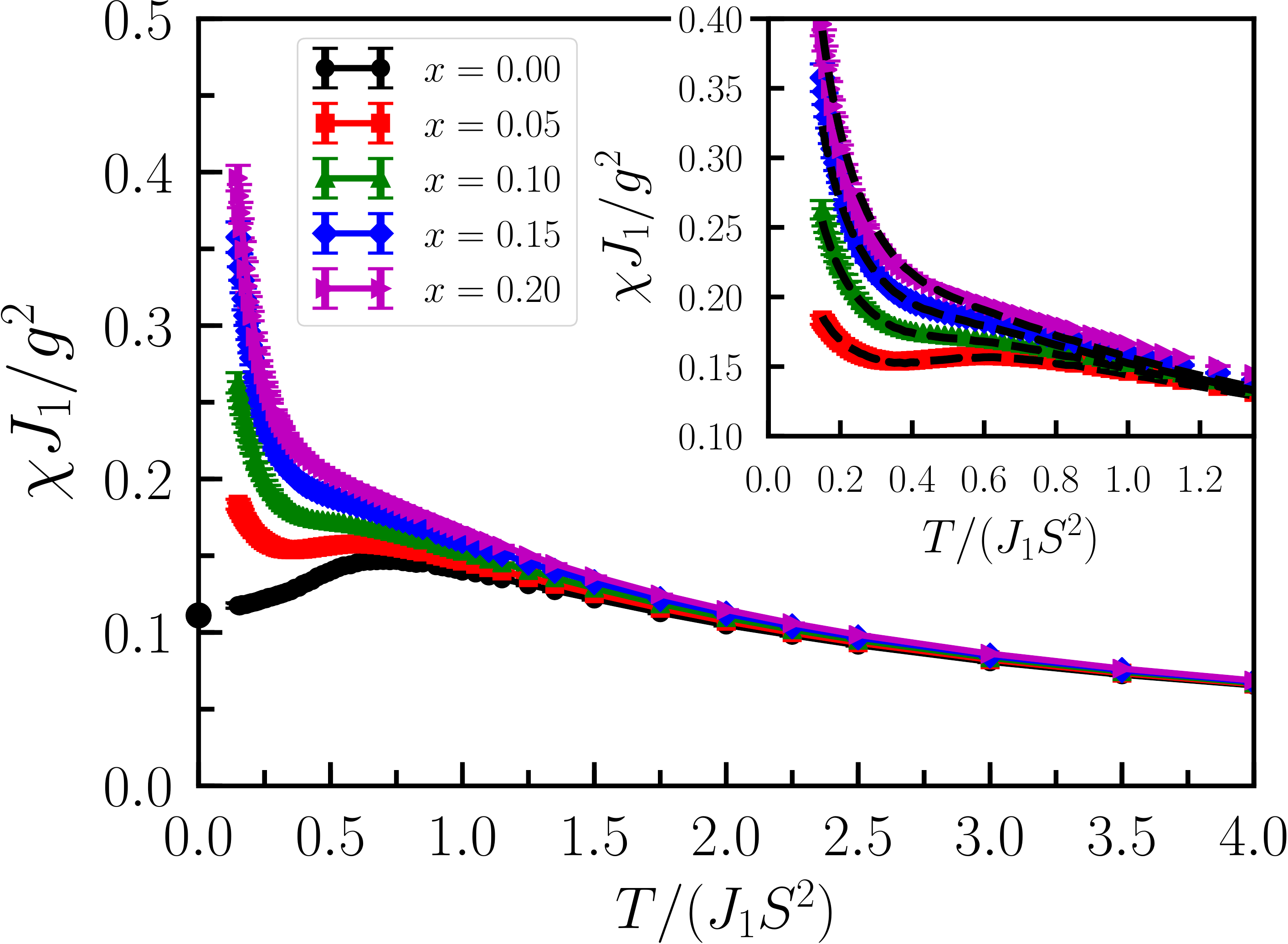}
\par\end{centering}
\caption{\label{fig:chi_heis}Uniform susceptibility $\chi$, in units of $g^2$, as function of temperature $T$ in the diluted Heisenberg model for different vacancy concentrations $x$, as obtained from classical MC simulations for $L=24$ and uniform $g$ factors.
The black dot shows the analytical result for $x=0$ in the low-temperature limit: $\chi(x=0)=g^2/(9J_1)$.
Inset: Zoom into the low-temperature region with the dashed lines showing the fit given by Eq.~\eqref{eq:chi_imp_fit}.
}
\end{figure}

As a warm up, let us consider the nearest-neighbor Heisenberg limit, $J_1 > 0$ and $K_1=\Gamma_1=J_{3}=0$.
We then have a N\'eel state as ground state, while long-range order is absent for finite $T>0$.
In the classical limit, the uniform susceptibility in the low-temperature limit is\citep{yosida} $ \chi(x=0)=g^2/(9J_1) + \mathcal O(T) $, which agrees with our numerical simulations, see black dot in Fig.~\ref{fig:chi_heis}.

As we turn on the dilution, the asymptotic Curie-Weiss temperature in Eqs.~\eqref{eq:cw_par} and \eqref{eq:cw_perp} is now reduced as $\Theta^{{\rm CW}}(x)=(1-x)\Theta^{{\rm CW}}(0)$, where $\Theta^{{\rm CW}}(0)=-cJ_1$ corresponds to the $x=0$ value, and $c = S^2$ [$c=S(S+1)$] for classical (quantum) spins. This is because we effectively diminish the number of neighbors of a given site.

Besides this overall suppression of the magnetic energy scale, we also observe a Curie-like tail in the low-temperature magnetic susceptibility, Fig.~\ref{fig:chi_heis}. We fit the disorder-averaged value of the susceptibility in this regime as
\begin{equation}
\chi(x)=g^2 \left[x\cdot\frac{m^{2}}{T}+\left(1-x\right)\cdot\chi(0)\right],\label{eq:chi_imp_fit}
\end{equation}
where $\chi(0)$ is the low-temperature susceptibility in the clean case, and we find that $m^{2}=AS^{2}/3$, with $A$ a numerical prefactor.

This result can be rationalized as follows: A single vacancy induces a local impurity moment because the magnetization of each sublattice no longer cancel. Because the bulk magnetic order is destroyed by thermal fluctuations, the direction of this impurity moment is not fixed but is free to rotate with the local orientation of the bulk magnetic domain
surrounding it, leading to a low-temperature Curie-like response at low fields.\cite{sbv99,sv03, hoglund03,sushkov03,eggert07,wav12}
Since our focus is the low-$T$ regime, for a finite concentration of impurities we are interested in the limit $\xi \gg l_{\rm imp}$, where $\xi$ is the bulk correlation length and $l_{\rm imp}\propto x^{-1/2}$ is the mean-impurity distance. In this case, the effective total impurity moment  $ S_{\rm eff} $ is given by the difference of the vacant sites on each sublattice: $S^2_{\rm eff} \sim x N S^2$ and we arrive at the Curie-tail contribution in Eq.~\eqref{eq:chi_imp_fit}. If we take $A \approx 2/3$,  Eq.~\eqref{eq:chi_imp_fit} fits the numerical results remarkably well for small $x$, see inset of Fig.~\ref{fig:chi_heis}. As we increase $x$, the mean impurity separation diminishes and corrections to the single-impurity result become apparent, although the leading effect is still well described by Eq.~\eqref{eq:chi_imp_fit}.


\subsection{Anisotropy evolution with dilution}

\begin{figure*}[t]

\textbf{Model 1} \kern 11pc \textbf{Model 2} \kern 10pc \textbf{Model 3}
\vspace{2mm}

\includegraphics[width=0.695\columnwidth]{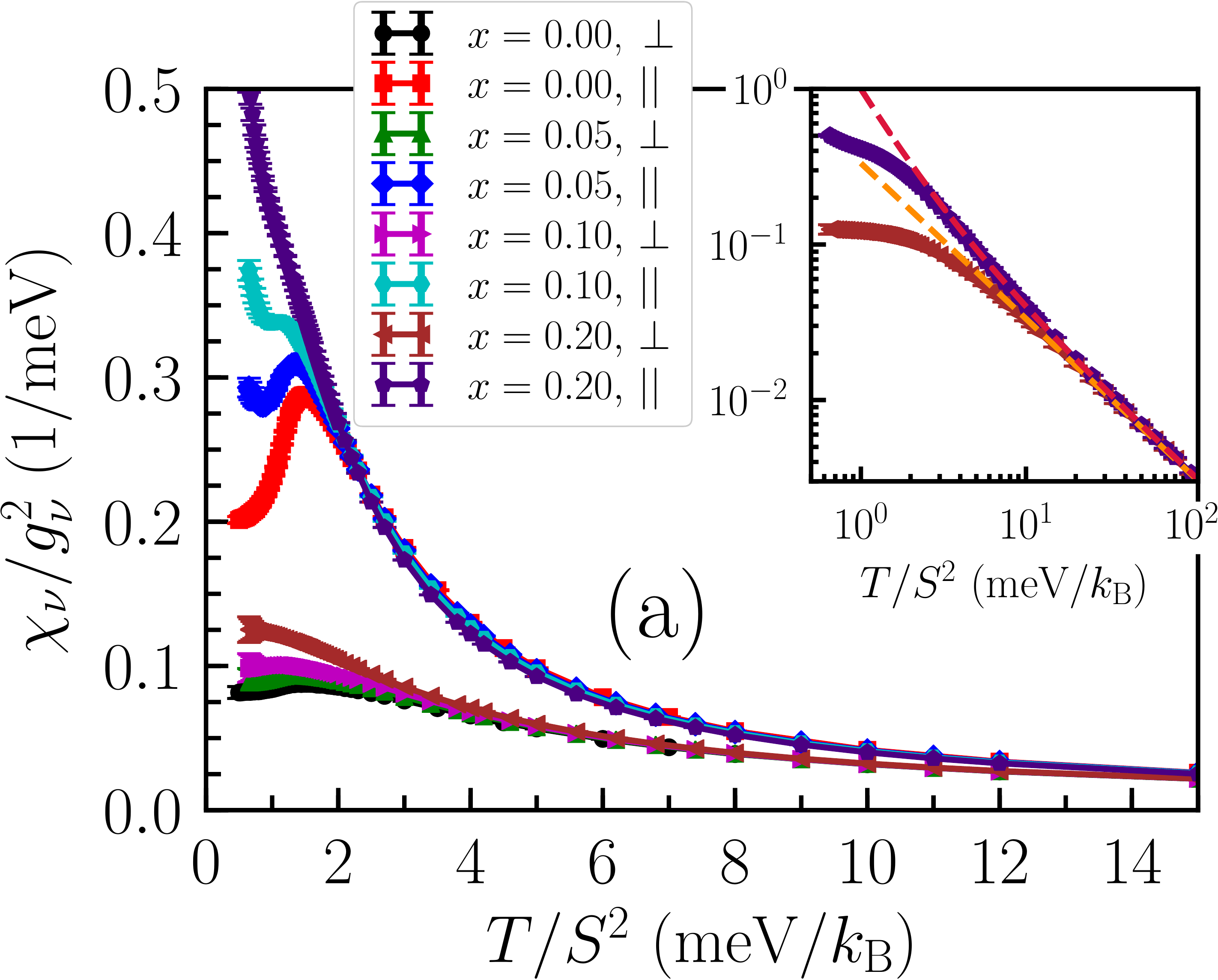}$\;$
\includegraphics[width=0.66\columnwidth]{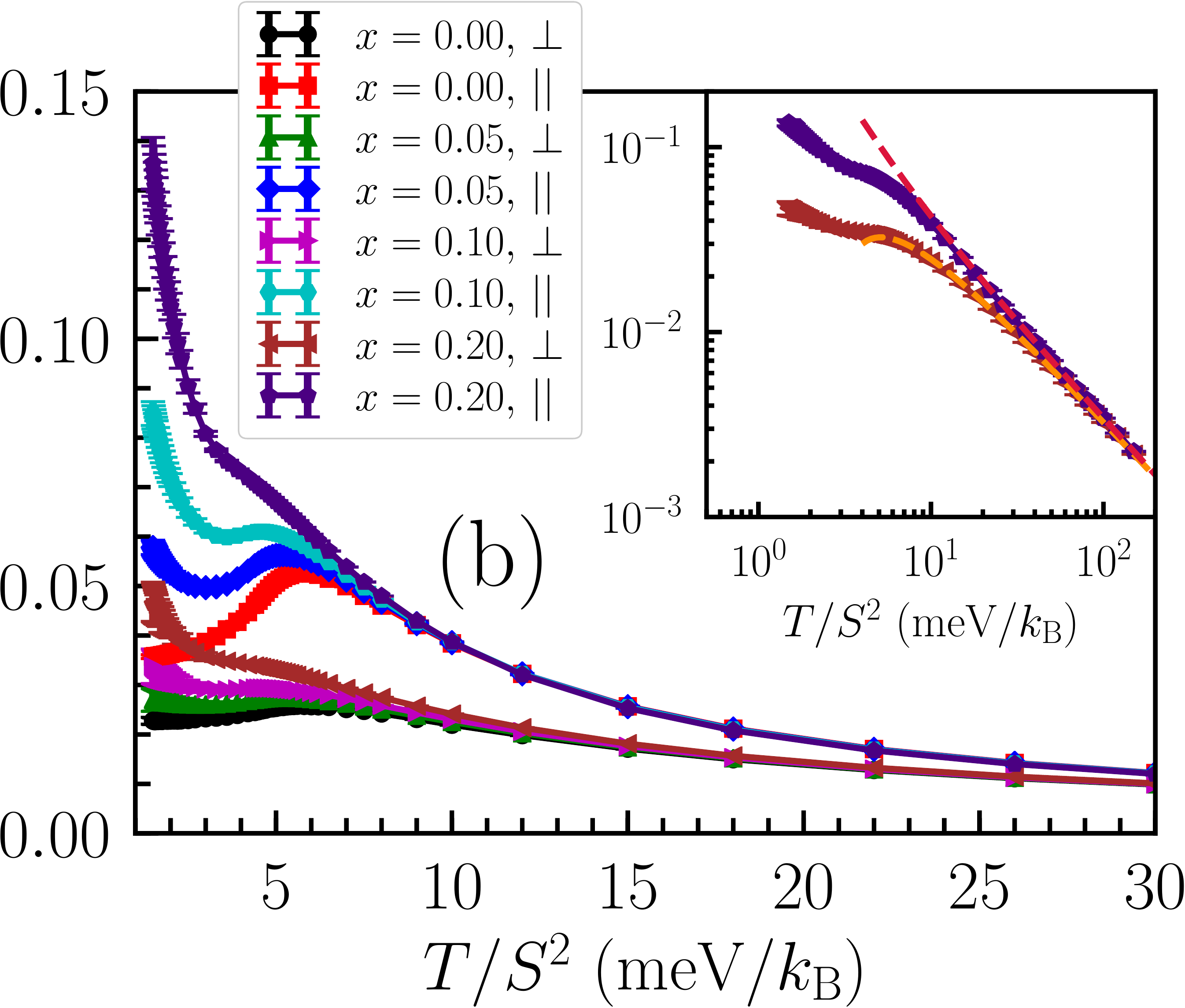}$\;$
\includegraphics[width=0.66\columnwidth]{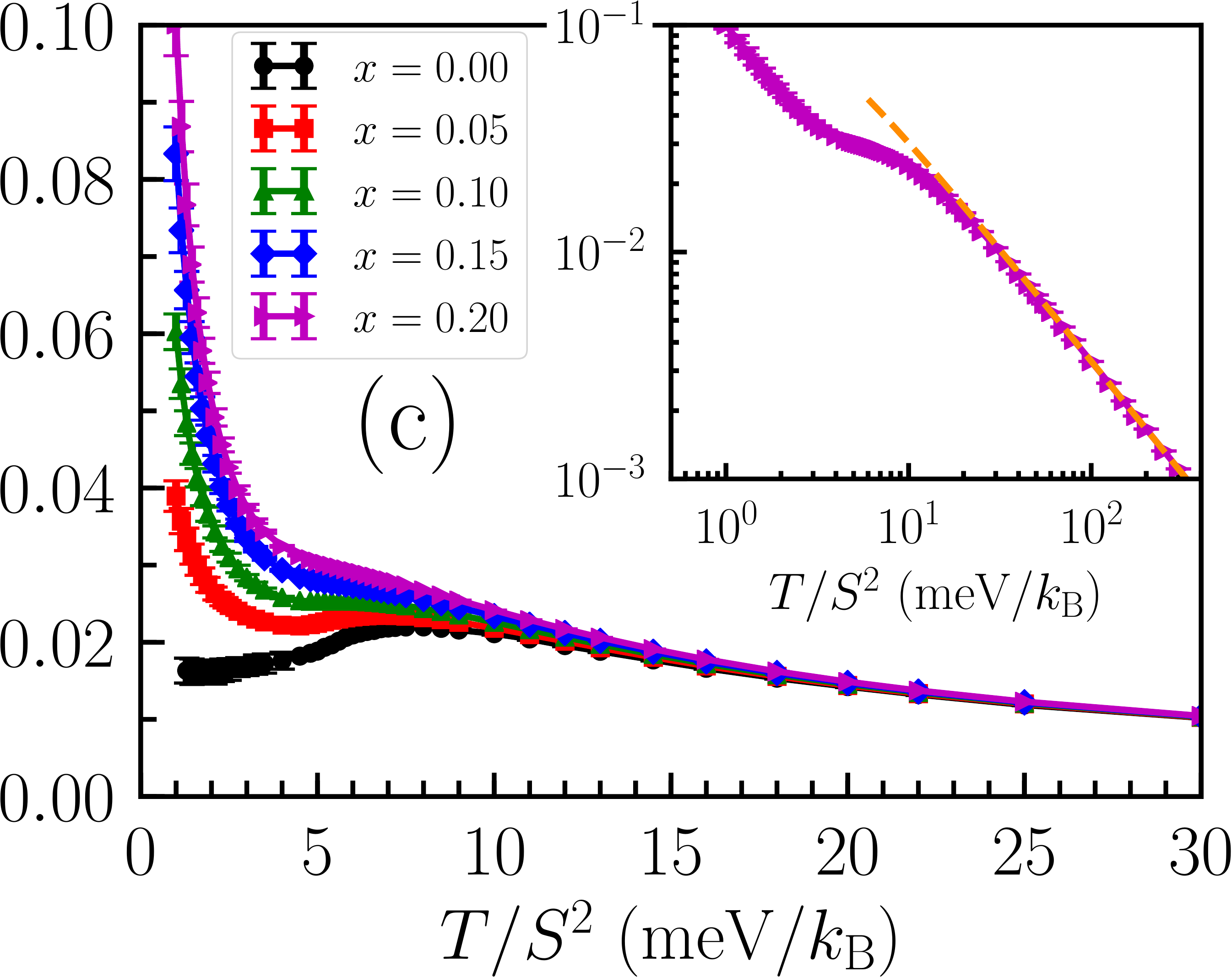}
\caption{\label{fig:chi_dilution}
Uniform susceptibilities $\chi_{\nu}$, $\nu=\perp,\,\parallel$, as function of temperature $T$ in units of $\gperp^2$ and $\gpar^2$, respectively, for Model 1 (a), Model 2 (b), and Model 3 (c), from classical MC simulations on an $L=24$ lattice.
Insets: Susceptibilities for $x=0.20$ in a log-log scale. The dashed lines correspond to the high-temperature results, Eqs.~\eqref{eq:cpar_highT} and \eqref{eq:cperp_highT}.
}
\end{figure*}

We now investigate how the intrinsic anisotropy of the HK$\Gamma$ model evolves
with random dilution. Here, the long-ranged zigzag magnetic order is destroyed for $x>0$ in favor of a spin glass in the classical limit, and we expect that the peaks in the susceptibility can no longer be associated with the onset of order, but instead mark a crossover regime below which spin fluctuations are suppressed.
Actually, in $d=2$, the spin-glass order takes place only\cite{maiorano18} at $T=0$. Therefore, we can, in principle, expect a finite-$T$ susceptibility that resembles that of the Heisenberg limit. Our numerical results indeed indicate the presence of a Curie tail in the susceptibility, both for $\cpar$ and $\cperp$. This is demonstrated in Fig.~\ref{fig:chi_dilution}, which shows the susceptibilities as function of temperature for Models 1, 2, and 3.

\begin{figure}[b]

\begin{centering}

\textbf{Model 1} \kern 6.0pc \textbf{Model 2}
\vspace{2mm}

\includegraphics[width=0.515\columnwidth]{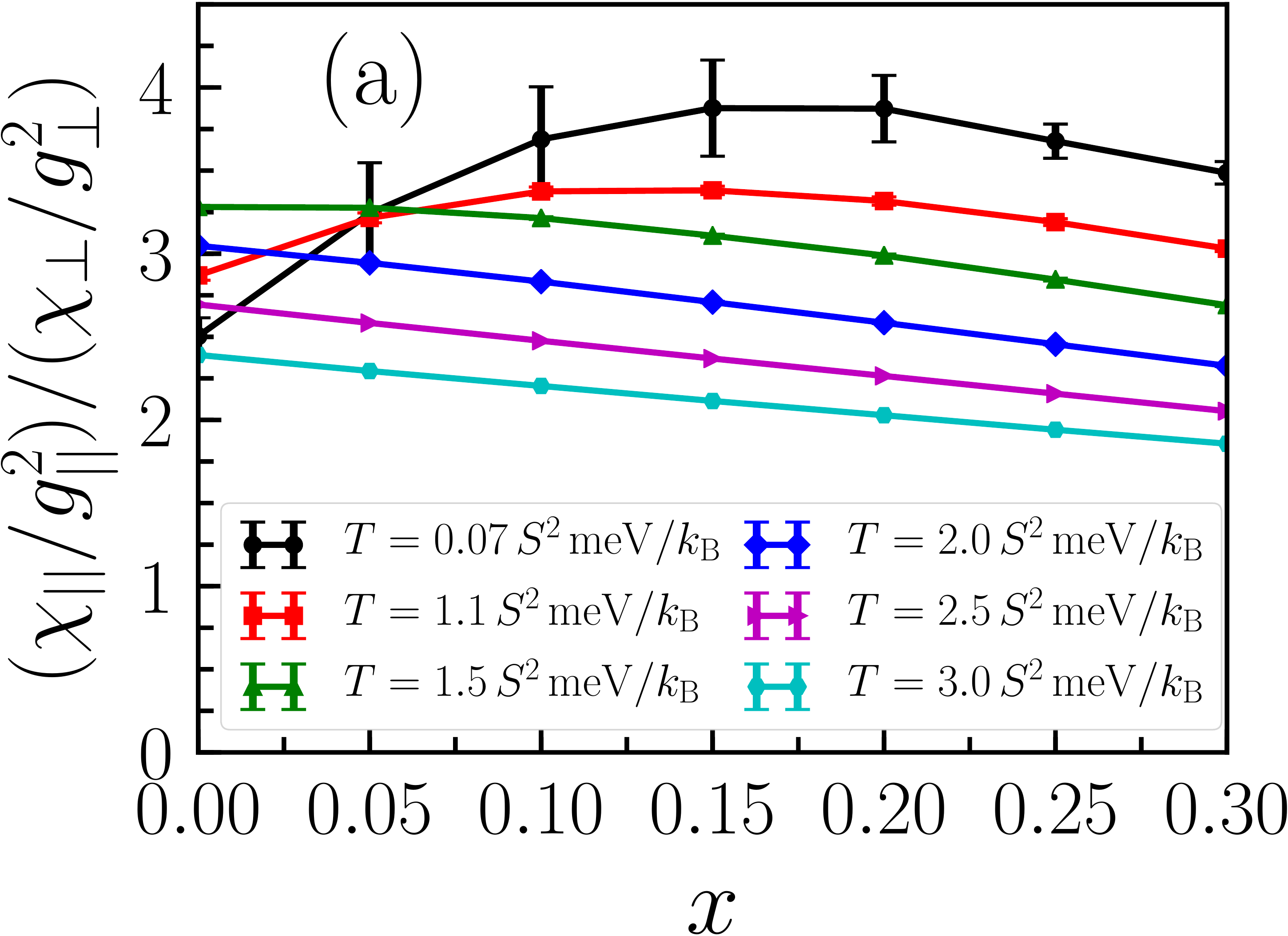}$\,$\includegraphics[width=0.485\columnwidth]{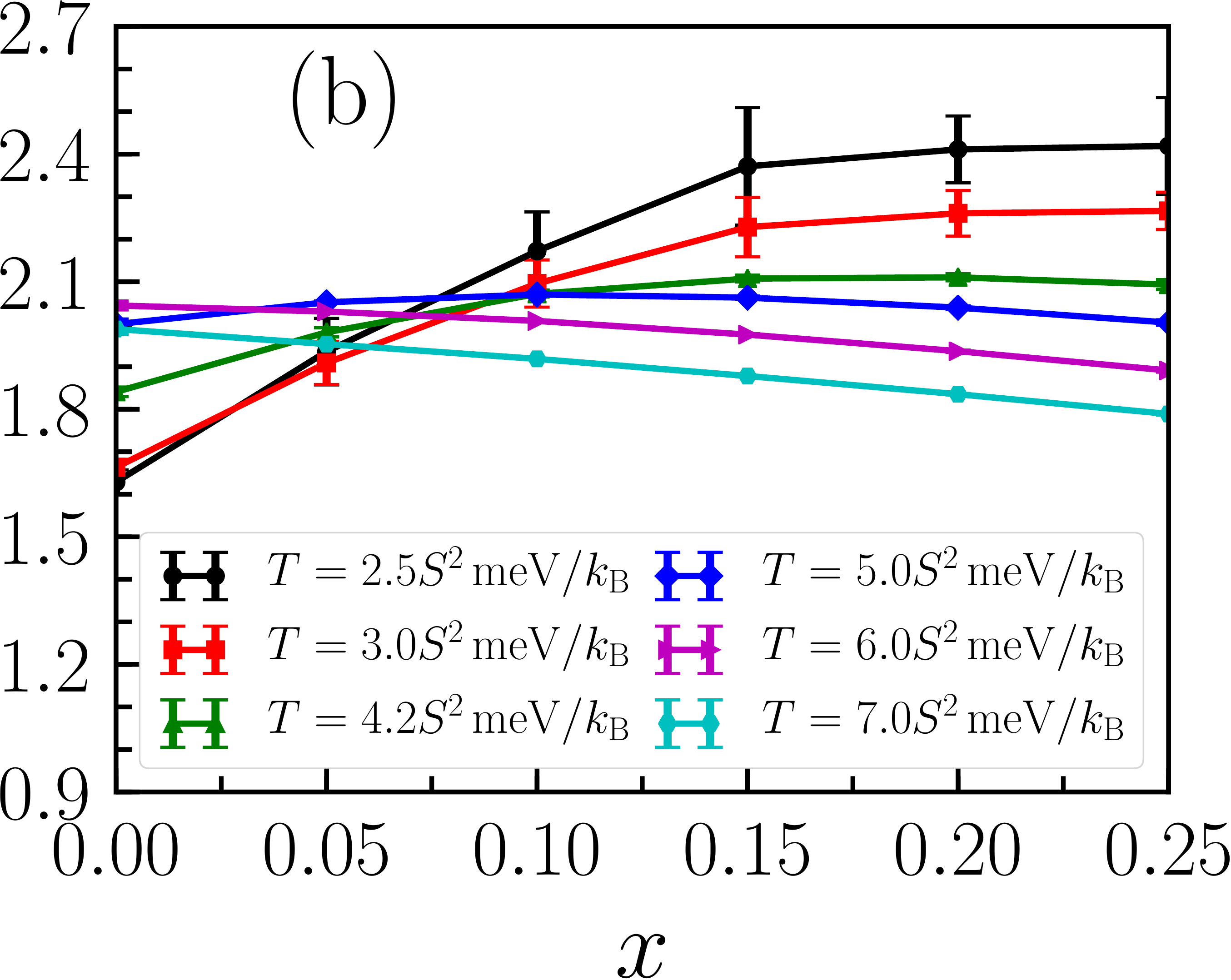}
\par\end{centering}
\caption{\label{fig:aniso_doping} Susceptibility anisotropy ratio $\cpar/\cperp$ in units of $(\gpar/\gperp)^2$ as function of doping level $x$ for different temperatures for Model 1 (a) and Model 2 (b), from classical MC simulations on an $L=24$ lattice.
For small doping level $x$, the low-temperature anisotropy increases with increasing $x$, while its high-temperature counterpart decreases.
}
\end{figure}

The evolution of the anisotropy ratio $\cpar/\cperp$ as function of doping $x$ for Models 1 and 2 is displayed in Fig.~\ref{fig:aniso_doping}. For small doping levels, the low-temperature anisotropy increases with increasing $x$, while at the same time high-temperature anisotropy decreases.
One way to rationalize this observation is to realize that the introduction of vacancies leads to an overall reduction of the exchange energy scales. Therefore, upon diluting, one effectively shifts the system to higher temperatures, which enhances the anisotropy at low temperatures, but suppresses it at high temperatures, cf.\ Fig.~\ref{fig:chi_clean}.
For coupled layers, this increase in the susceptibility upon cooling should be bounded at low $T$ by the freezing temperature $T_\mathrm{g}$, and we expect this bound to be suppressed with $x$.\cite{hk_depl14}

Combining Eqs.~\eqref{eq:cpar_highT} and ~\eqref{eq:cperp_highT} with the dilution induced suppression of the Curie-Weiss temperature, $\Theta_{\parallel,\perp}^{{\rm CW}}(x) = (1-x)\Theta_{\parallel,\perp}^{{\rm CW}}(0)$, where $\Theta_{\parallel,\perp}^{{\rm CW}}(0)$ is their clean value, we obtain the high temperature disorder evolution of the susceptibilities %
\begin{align}
\chi_{\parallel,\perp}(x)=\chi_{\parallel,\perp}(0) - x g_{\parallel,\perp}^2\frac{c}{3T^2}\Theta_{\parallel,\perp}^{{\rm CW}}(0)+\mathcal{O}(T^{-3}),\label{eq:aniso_highT}
\end{align}
where $\chi_{\parallel,\perp}(0)$ are the $x=0$ susceptibilities. As shown in the insets of Fig.~\ref{fig:chi_dilution}, this equation is consistent with our numerical data.
Therefore, whether disorder enhances or suppresses the high-temperature susceptibility can be traced back to the sign of the respective clean Curie-Weiss temperature.
Consequently, at high temperatures, the in-plane susceptibility decreases with increasing $x$ in Models 1 and 2, while the out-of-plane susceptibility increases, at least in Model 2. This trend agrees with our simulation results, cf.\ Fig.~\ref{fig:aniso_doping}.
For Model 3, where the anisotropy is absent, we have $\Theta_{0}^{{\rm CW}}<0$, and we obtain a corresponding increase in $\chi(x)$ for increasing $x$.

\subsection{\label{sec:experiment}Comparison to experiments}

\subsubsection{Na$_2$(Ir$_{1-x}$Ti$_x$)O$_3$}

The diluted version of Model 3 may be relevant for Na$_2$(Ir$_{1-x}$Ti$_x$)O$_3$. In fact, our results shown in Fig.~\ref{fig:chi_dilution}(c) are broadly consistent with the reported measurements in this compound:\cite{manni14} (i) Upon the introduction of vacancies, the long-range order gives way to a spin glass. This glass state is identified via the separation between field-cooled and zero-field-cooled susceptibilities as measured at a very low field of 5\,mT. (ii) Both the freezing temperature and the absolute value of the Curie-Weiss temperatures decrease with $x$.\cite{manni14,hk_depl14} (iii) The low-temperature part of the susceptibility is enhanced with $x$. However, the available experimental data do not address the anisotropy evolution with doping.

\subsubsection{$\alpha$-(Ru$_{1-x}$Ir$_x)$Cl$_3$}

For $\alpha$-(Ru$_{1-x}$Ir$_x)$Cl$_3$, the introduction of vacancies is accompanied by the return of two transition temperatures,\cite{kelley17,do18} indicating the presence of stacking faults in the doped samples. These two ordering temperatures, as well as the temperature of the structural first-order phase transition, decrease with $x$. Despite the broadening of the specific-heat peak with $x$, the experimental susceptibilities show no sign of separation between field-cooled and zero-field-cooled runs. However, these susceptibility measurements were performed at comparatively high fields $\geq 100$\,mT.\cite{kelley17,do18} A weakly glassy state can therefore not be fully excluded.

The behavior of $\cpar$ in Model 1 as displayed in Fig.~\ref{fig:chi_dilution}(a) describes qualitatively well the experimental results for $\alpha$-(Ru$_{1-x}$Ir$_x)$Cl$_3$: The low-temperature increase of the in-plane susceptibility with $x$ can be linked to the impurity-induced Curie tail, whereas its decrease at high temperature can be associated with the suppression of $\tcwpar$ with doping. On the other hand, the evolution of $\cperp$ with doping in $\alpha$-(Ru$_{1-x}$Ir$_x)$Cl$_3$ is more involved. Equation~\eqref{eq:aniso_highT} suggests a slight increase of $\cperp$ with doping at high temperatures, opposite to the experimental observation. Since this is a weak effect, we suspect again that the structural transition may be relevant, especially because it affects $\cperp$ more strongly. The non-monotonic behavior of $\cperp$ as function of $x$ at low temperatures is also not captured by our classical models: The MC results show a monotonic increase of $\cperp$ with $x$, although it does so more slowly than $\cpar$.

Overall, we believe one should be very careful when comparing susceptibilities between different samples. This is because the direction-specific susceptibilities are extremely sensitive to stacking faults, which are inevitably introduced with doping.\cite{kelley17} Typically, the in-plane susceptibility decreases upon the introduction of stacking faults, while at the same time the out-of-plane susceptibility increases. Lower-quality samples are therefore more isotropic than higher-quality samples.\cite{majumder15,sears15,kelley18a} One observable that is potentially less sample dependent is the averaged (``powder'') susceptibility $\chi_{\rm{avg}} = (2\cperp + \cpar)/3$, which, however, naturally lacks information about the anisotropy.


\section{\label{sec:conclusion}Conclusions and outlook}

We have studied the magnetic susceptibilities in extended Heisenberg-Kitaev models, with focus on their spatial anisotropy and its evolution with temperature and magnetic dilution. We have reported results from both high-temperature expansion and large-scale classical MC simulations.

For \rucl, our findings reconcile different results reported in the literature. We show that a set of exchange parameters, originally obtained inside the ordered phase at low temperature, describes well the behavior of the susceptibilities and their anisotropy over a large range of temperatures. This set includes a sizeable off-diagonal $\Gamma_1$ interaction and a moderate $g$-factor anisotropy, with the temperature dependence of the anisotropy arising solely from the $\Gamma_1$ interaction.

In \nio, the susceptibility anisotropy is mostly temperature independent, which is consistent with the assumption of a small $\Gamma_1$. The experimental behavior is well captured by our model, if one assumes a suitable $g$-factor anisotropy.

In the presence of doping, we find that the high-temperature susceptibilities may be mildly enhanced or suppressed depending on the sign of the clean Curie-Weiss temperature. At low temperatures, our model generically predicts the suppression of long-range order and the emergence of a spin-glass state, which agrees with the experimental observation for Na$_2$(Ir$_{1-x}$Ti$_x$)O$_3$.\cite{manni14} By contrast, the experiments on $\alpha$-(Ru$_{1-x}$Ir$_x)$Cl$_3$ have been interpreted in terms of a disordered spin liquid,\cite{kelley17,do18,do20,baek20} which goes beyond our semiclassical modeling. At low temperatures, we also find a Curie-tail contribution to the magnetic susceptibility, which gives rise not only an enhancement of the individual susceptibilities, but also of their anisotropy. This is in accordance with the experimental trends observed at small doping.

For $\alpha$-(Ru$_{1-x}$Ir$_x)$Cl$_3$, the qualitative behavior of in-plane susceptibility $\cpar$ is well captured by our modeling. By contrast, the agreement for the out-of-plane susceptibility $\cperp$ is less satisfactory.
This may be linked to the presence of stacking faults, which appears to have a particularly strong influence on $\cperp$.
In this respect, a systematic study of the out-of-plane susceptibility $\cperp$ on different \rucl\ and $\alpha$-(Ru$_{1-x}$Ir$_x)$Cl$_3$ samples of varying quality would be desirable.
For Na$_2$(Ir$_{1-x}$Ti$_x$)O$_3$, we are not aware of any measurements on the anisotropy evolution, but our results for the averaged susceptibility agree well with the experiments.\cite{manni14}

Ideally, the evolution of the magnetic anisotropy in Kitaev materials, both with temperature and doping, could be used to further constrain the values of the symmetry allowed exchange constants, in order to complement the usual modeling in the low-temperature regime. This, however, requires more information concerning the sample dependence of the experimental behavior seen in the current data.

On the theoretical side, we have employed in this work a simple semiclassical modeling. While this approximation should yield reasonable results at intermediate and high temperatures, qualitatively new physics can emerge in the low-temperature quantum regime. This is in particular true when the ground state realizes a (disordered) Kitaev quantum spin liquid.\cite{zschocke15, knolle19a}
While the low-doping behavior may potentially be understood in terms of isolated vacancies,\cite{willans10,willans11} a description of the behavior at finite doping and its temperature dependence requires a full many-body quantum computation.\cite{nasu15,kimchi18, liu18b}
This represents an excellent direction for future research.


\begin{acknowledgments}
We thank B. B\"uchner, P. C\^onsoli, S. Nagler, S. Rachel, and A. U. B. Wolter for discussions and collaborations on related topics.
ECA was supported by CNPq (Brazil) Grants No.\ 406399/2018-2 and No.\ 302994/2019-0, and FAPESP (Brazil) Grant No.\ 2019/17026-9.
The work of LJ is funded by the Deutsche Forschungsgemeinschaft (DFG) through the Emmy Noether program (JA 2306/4-1, project id 411750675).
LJ and MV acknowledge support by the DFG through SFB 1143 (project id 247310070) and the W\"urzburg-Dresden Cluster of Excellence on Complexity and Topology in Quantum Matter---\textit{ct.qmat} (EXC 2147, project id 390858490).
ECA acknowledges the hospitality of TU Dresden, where part of this work was performed.
\end{acknowledgments}


\appendix

\section{\label{sec:gprime}Trigonal distortions and $\Gamma_1^{\prime}$}

\begin{figure}[t]
\begin{centering}

\textbf{HK$\Gamma\Gamma'$ model}
\vspace{2mm}

\includegraphics[width=0.96\columnwidth]{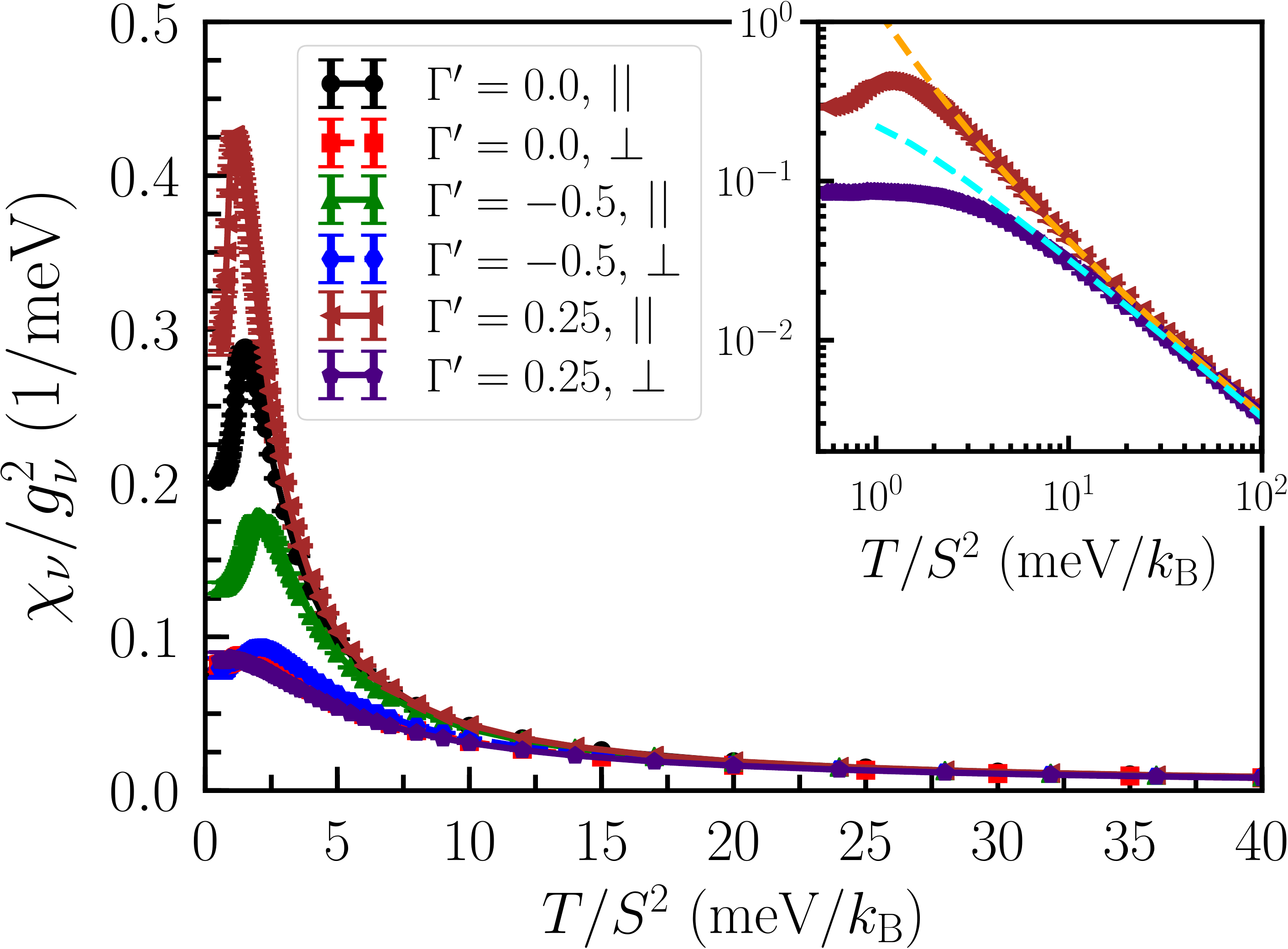}
\par\end{centering}
\caption{\label{fig:chi_gprime}In-plane, $\nu=\parallel$, and out-of-plane, $\nu=\perp$, uniform magnetic
susceptibilities $\chi_{\nu}$ in units of $\gpar^2$ and $\gperp^2$, respectively, as function of temperature for different values of $\Gamma^{\prime}_1$ (in units of meV). These results were obtained from classical MC simulations on an $L=24$ lattice with $J_1$, $K_1$, $\Gamma_1$, and $J_3$ chosen as in Model 1.
Inset: Susceptibilities for $\Gamma^{\prime}_1=0.25$\,meV in a log-log scale. The dashed lines show the results from the high-temperature expansion.
}
\end{figure}

The Hamiltonian presented in Eq.~\eqref{eq:hkg} does not contain
all nearest-neighbor spin interactions that are allowed by the $C_{3}^*$ symmetry. This symmetry corresponds to a $2\pi/3$
spin rotation about the $\left[111\right]$ direction in spin space
combined with a $2\pi/3$ lattice rotation about one site. A trigonal
distortion (compression or elongation along the $\left[111\right]$
axis) fully preserves this symmetry and generates an extra nearest-neighbor
magnetic coupling\cite{rau14b}
\begin{equation}
\mathcal{H}^{\prime}=\Gamma_1^{\prime}\sum_{\left\langle ij\right\rangle _{\gamma}}\left(S_{i}^{\alpha}S_{j}^{\gamma}+S_{i}^{\beta}S_{j}^{\gamma}+S_{i}^{\gamma}S_{j}^{\alpha}+S_{i}^{\gamma}S_{j}^{\beta}\right),\label{eq:hgprime}
\end{equation}
using the same notation as in Eq.~\eqref{eq:hkg}. For \rucl, ab initio calculations suggest a small negative $\Gamma^{\prime}_1$;\citep{kim16,winter16,eichstaedt19} however, recently it has been argued that empirical constraints require a sizeable positive $\Gamma^{\prime}_1$.\citep{maksimov20}
We have simulated $\mathcal{H}+\mathcal{H}^{\prime}$ in the high-temperature regime to investigate the effects of $\Gamma^{\prime}_1$ on the magnetic anisotropy.

In Fig.~\ref{fig:chi_gprime}, we show the in-plane and out-of-plane
magnetic susceptibilities for different values of $\Gamma^{\prime}_1$.
In the clean limit, we see that the anisotropy ratio $\cpar/\cperp$ decreases (increases) for $\Gamma_1^{\prime}<0$ ($\Gamma_1^{\prime}>0$), resulting from a strong dependence of the in-plane susceptibility $\cpar$ on $\Gamma_1^{\prime}$ at low temperatures.
Note that when $J_1$, $K_1$, $\Gamma_1$, and $J_3$ are chosen as in Model 1, the zigzag ground state is stable only up to $\Gamma^{\prime}_1 \lesssim 0.45$\,meV.
For large negative $\Gamma'_1 \simeq -\Gamma_1$, the anisotropy inverts, such that $\cpar$ drops below $\cperp$ in the low-temperature limit. We have checked that dilution does not alter this trend.

Inclusion of $\Gamma^{\prime}_1$ in the high-temperature expansion modifies the Curie-Weiss temperatures accordingly to
\begin{align}
\tcwpar & = -\frac{c}{3}\left[3\left(J_{1}+J_{3}\right)+K_1-\left(\Gamma_1 + 2\Gamma_1^{\prime}\right) \right],\label{eq:cw_par_gp}\\
\tcwperp & = -\frac{c}{3}\left[3\left(J_{1}+J_{3}\right)+K_1+2(\Gamma_1+2\Gamma_1^{\prime})\right],\label{eq:cw_perp_gp}
\end{align}
and we show in the inset of Fig.~\ref{fig:chi_gprime} that this expression describes well the MC results at elevated temperatures.

Note that $\Gamma^{\prime}_1$ has a stronger influence on the low-temperature part of the anisotropy than on its high-temperature tail. This indicates that inclusion of a small positive $\Gamma^{\prime}_1$ may help to improve agreement with the experimental results for \rucl.
For instance, choosing $\Gamma^{\prime}_1 = 0.25$\,meV and $J_1$, $K_1$, $\Gamma_1$, and $J_3$ as in Model 1, together with a slightly reduced $g$-factor anisotropy of $\gpar/\gperp \simeq 1.33$, we obtain $\cpar/\cperp \approx 1.9$, $8.8$, $6.2$ above room temperature, near the transition, and in the low-temperature limit, respectively. This is in well agreement with the experimental values for \rucl.


%


\begin{thebibliography}{85}
\bibitem [{inp()}]{inplanefoot}%
  \BibitemOpen
  \href@noop {} {}\bibinfo {note} {For small fields, i.e., in the
  linear-response limit, the susceptibility is identical for all in-plane
  directions provided that the three honeycomb bonds are
  equivalent.}\BibitemShut {Stop}%
\bibitem [{nio()}]{niosuscfoot}%
  \BibitemOpen
  \href@noop {} {}\bibinfo {note} {In Ref.~\onlinecite{singh10}, the
  susceptibilities along the crystallographic $c$ axis and an axis
  perpendicular to $c$ have been measured, which in the $C2/m$ structure of
  \nio\ only approximately agrees with axes perpendicular and parallel to the
  honeycomb plane.}\BibitemShut {Stop}%
\end{thebibliography}
\end{document}